\documentclass[conference]{IEEEtran}
\IEEEoverridecommandlockouts
\usepackage{cite}
\usepackage{amsmath,amssymb,amsfonts}
\usepackage{algorithmic}
\usepackage{graphicx}
\usepackage{textcomp}
\usepackage{xcolor}
\usepackage[utf8]{inputenc}
\usepackage{array}
\usepackage{wrapfig}
\usepackage{multirow}
\usepackage{tabu}
\def\BibTeX{{\rm B\kern-.05em{\sc i\kern-.025em b}\kern-.08em
    T\kern-.1667em\lower.7ex\hbox{E}\kern-.125emX}}
\begin{document}

\title{Technical Debt: Identify, Measure and Monitor}

\author{\IEEEauthorblockN{Nikhil Oswal}
\IEEEauthorblockA{\textit{School of Electrical Engineering and Computer Science (EECS)
} \\
\textit{University of Ottawa}\\
Ottawa, Canada \\
noswa023@uottawa.ca}

}

\maketitle

\begin{abstract}

Technical Debt is a term begat by Ward Cunningham to signify the measure of adjust required to put a software into that state which it ought to have had from the earliest starting point. Often organizations need to support continuous and
fast delivery of customer value both in short and a long-term
perspective and later have to compromise with
the quality and productivity of the software. So, a simple
solution could be to repay the debts as and when they are
encountered to avoid maintainability cost and subsequent
delays. Therefore, it has become inevitable to identify and come up with techniques so as to know when, what
and how TD items to repay. This study aims to explore on how to identify, measure and monitor technical debt using SonarQube and PMD.

\end{abstract}


\section{Introduction}
The notion of ``technical debt'' was coined by Ward Cunningham at the OOPSLA conference in 1992. The original meaning as used by Cunningham was ``all the not quite right code which we postpone making it right.'' \cite{b1}. With this statement he was referring to the inner quality of the code. Later the term was extended to imply all that should belong to a properly developed software system, but which was purposely left out to remain in time or in budget, system features such as error handling routines, exception conditions, security checks and backup and recovery procedures and essential documents such as the architecture design, the user guide, the data model and the updated requirement specification. All of the many security and emergency features can be left out by the developer and the users will never notice it until a problem comes up. It is however very important to deal with this left debts. Here, technical debt management
(TDM) comes into picture which involves various processes and
tools to identify, represent, measure, prioritize and prevent
Technical debt \cite{b2}.

This paper focuses on conducting these
aforementioned technical management activities mainly-
Identification, Representation, Estimation, Monitoring,
Repayment and Prevention on two projects - Core Java 8 and Booking Manager using tools
- SonarQube and PMD. Through these tools,
the paper analyses projects in depth and extracts out
all the possible forms of technical debt, calculates the estimation
effort required to fix it, tries to monitor evolution of debt with
time using appropriate technique and even proposes ways of
repaying and preventing the debt in limited time period.
The two projects which we have taken for analysis have been
described as follows. \\
\textbf{Project 1: Core Java 8} - Java and XML based project with a total of 1.9k lines of code; Table I depicts the details with technology Stack.

 \begin{table}[hbt!]
\caption{Project 1} % title name of the table
\centering % centering table
\begin{tabular}{l c c rrrrrrr} % creating 10 columns
\hline % inserting double-line
 Java LOC & 1.7K 
\\ 
\hline % inserts single-line
 XML LOC & 200
\\ [0.5ex]
\hline % inserts single-line
 Total Lines of Code & 1948
\\ [0.5ex]
\hline % inserts single-line
 Lines & 2594
\\ [0.5ex]
\hline % inserts single-line
 Statements & 509
\\ [0.5ex]
\hline % inserts single-line
 Functions & 354
\\ 
\hline % inserts single-line
 Classes & 93
\\ 
\hline % inserts single-line

\hline % inserts single-line
\end{tabular}
\label{tab:PPer}
\end{table}
\noindent \textbf{Project 2: Booking Manager} - A web application with a total of 70k lines of code; Table II depicts the details with technology Stack.

\begin{table}[hbt!]
\caption{Project 2} % title name of the table
\centering % centering table
\begin{tabular}{l c c rrrrrrr} % creating 10 columns

\hline % inserting double-line
 JavaScript LOC & 41K 
\\ 
\hline % inserts single-line
 CSS LOC & 17K
\\ [0.5ex]
\hline % inserts single-line
 JSP LOC & 4.9K
\\ [0.5ex]
\hline % inserts single-line
 Java LOC & 4.5K
\\ [0.5ex]
\hline % inserts single-line
 HTML LOC & 2.9K
\\ [0.5ex]
\hline % inserts single-line
 XML LOC & 350
\\ [0.5ex]
\hline % inserts single-line
 Total Lines of Code & 70K
\\ [0.5ex]
\hline % inserts single-line
 Lines & 94417
\\ [0.5ex]
\hline % inserts single-line
 Statements & 24646
\\ [0.5ex]
\hline % inserts single-line
 Functions & 3703 
\\ 
\hline % inserts single-line
 Classes & 66
\\ 
\hline % inserts single-line

\hline % inserts single-line
\end{tabular}
\label{tab:PPer}
\end{table}

\noindent The remainder of the paper is structured as follows. Section
2 discusses about the entire Technical Debt Management activities.
In Section 3, we introduce a cost model for
estimating technical debt principal and also cover a new tool for
managing the debt followed by conclusion.

\section{MANAGING TECHNICAL DEBT USING TOOLS}

\subsection{Quality Assessment} 
\noindent Len Bass defines Quality Attributes [QA] as measurable or testable property of a system that is used to indicate how well the system satisfies the needs of its stakeholders. They mainly adhere to non-functional requirements. 

\noindent General quality attributes include Correctness, Reliability,
Adequacy, Learnability, Robustness, Maintainability,
Readability, Extensibility, Testability, Efficiency and
Portability. SonarQube allows accessing of three main quality
attributes - Reliability, Maintainability and Security.

We assess the above-mentioned
attributes using SonarQube tool as follows:
\begin{itemize}
\item
Reliability: It is measured as the probability of a system being
fully functional for a specified period of time without fail \cite{b3}.
\item
Maintainability: It measures how much capable the system is to
bring any kind of change with ease. The change can be due to
change of requirements, fixing of errors or implementation of
new features \cite{b3}.
\item
Security: It measures the capability of the system to withstand
any sort of malicious actions and prevent loss of information
\cite{b3}.
\end{itemize}
\noindent \textbf{Project 1: Core Java 8}
\begin{figure}[hbt!]
\centering
    \frame{\includegraphics[width=\columnwidth]{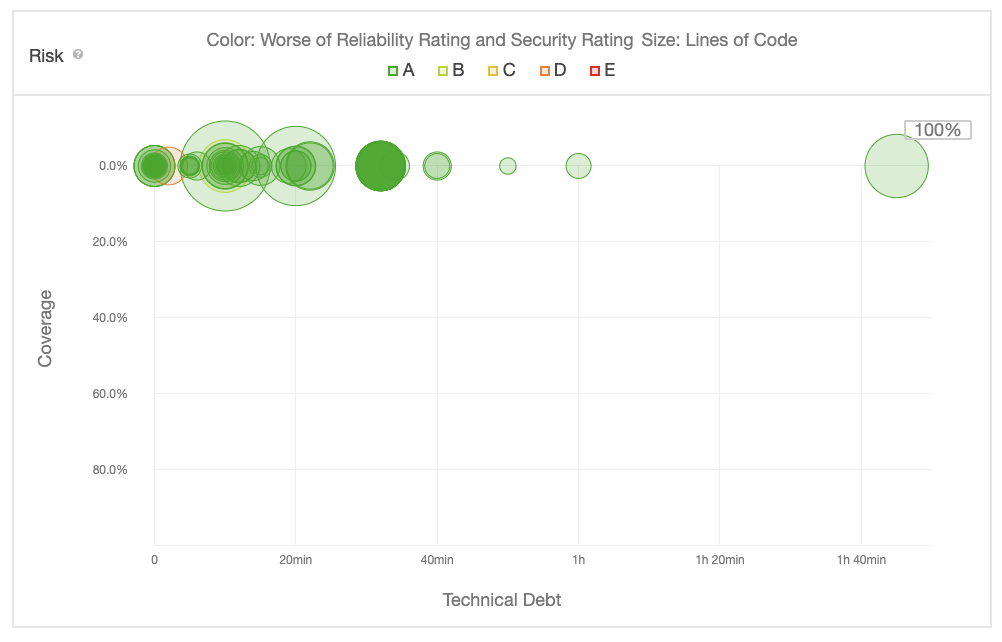}}
    \caption{Project Overview Using SonarQube.}
\end{figure}
\begin{itemize}
\item
Reliability: Using SonarQube, it has been observed that there is at least one
critical bug and overall, there are 2 bugs. It would take
around 10 minutes (estimated time) to fix these reliability issues
(shown in Fig. 2)\cite{b4}.
\begin{figure}[hbt!]
\centering
    \frame{\includegraphics[width=\columnwidth]{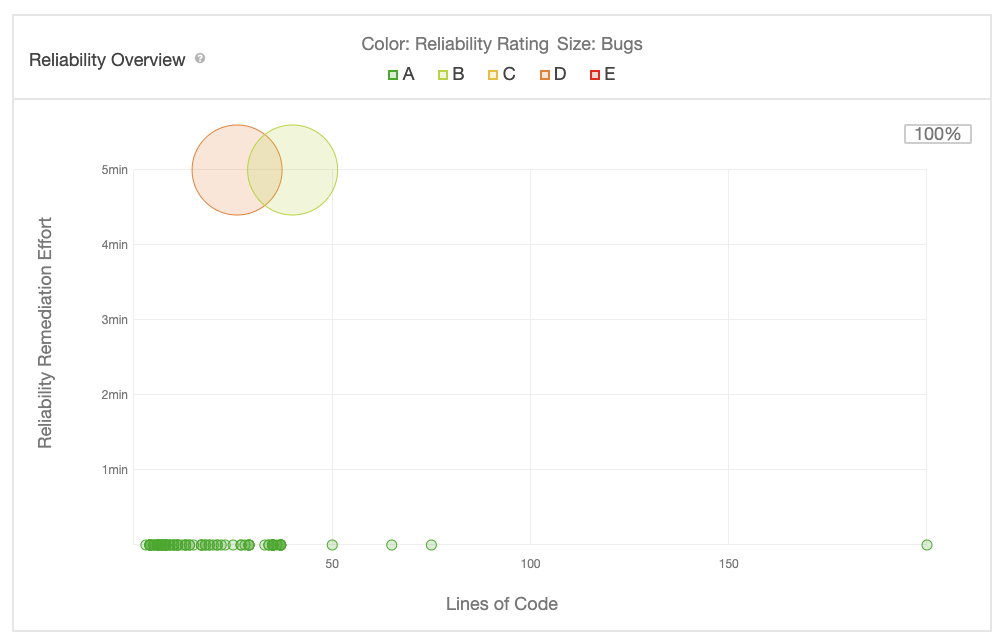}}
    \caption{Reliability Overview.} 
\end{figure}
\\
Here the red coloured bubble indicates critical bug and green one indicates minor bug.
\item
Security: With respect to Security, there are  3 vulnerabilities
observed out of which there one is a blocker
vulnerability which can make the whole application unstable
during production. It would take 45 minutes to fix all the vulnerability issues (depicted in Fig. 3)\cite{b4}.
\begin{figure}[hbt!]
\centering
    \frame{\includegraphics[width=\columnwidth]{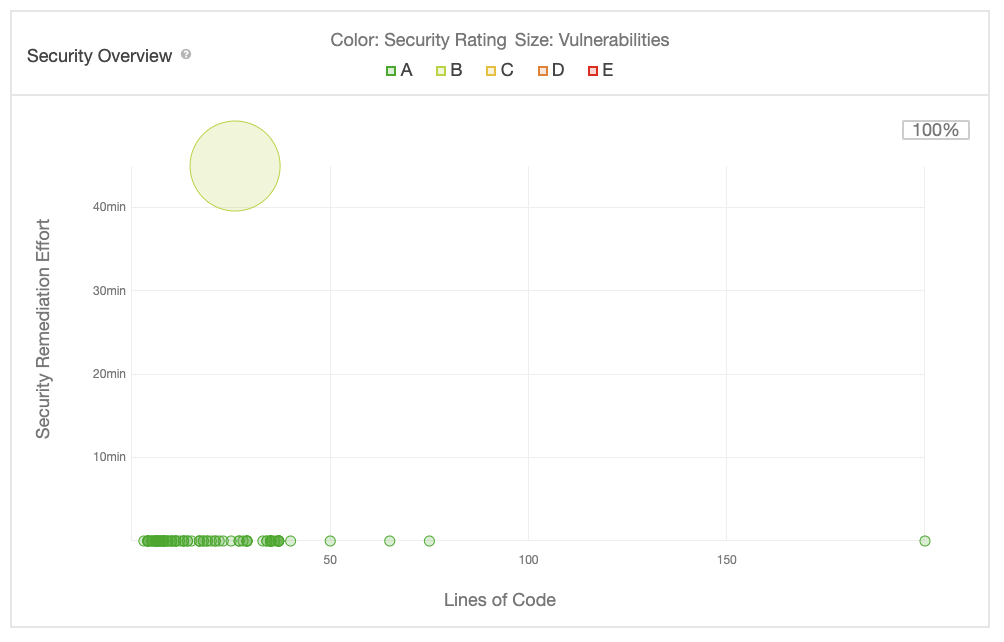}}
    \caption{Security Overview.} 
\end{figure}
\item
Maintainability: As far as Maintainability is concerned, there are 148 smells with ratio between the cost to develop the software and
the cost to fix it (i.e., the technical debt ratio) is 2.3\% as
depicted in Fig. 4. It would take 2 days and 6 hours to fix all the debts \cite{b4}.

\begin{figure}[hbt!]
\centering
    \frame{\includegraphics[width=\columnwidth]{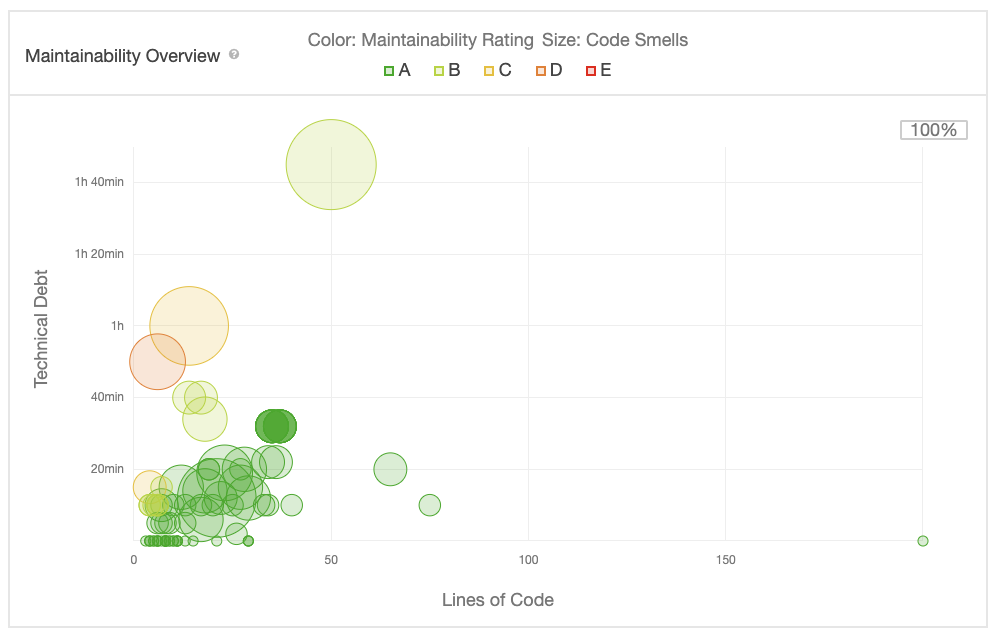}}
    \caption{Maintainability Overview.} 
\end{figure}

\end{itemize}

\noindent \textbf{Project 2: Booking Manager}
\begin{figure}[hbt!]
\centering
    \frame{\includegraphics[width=\columnwidth]{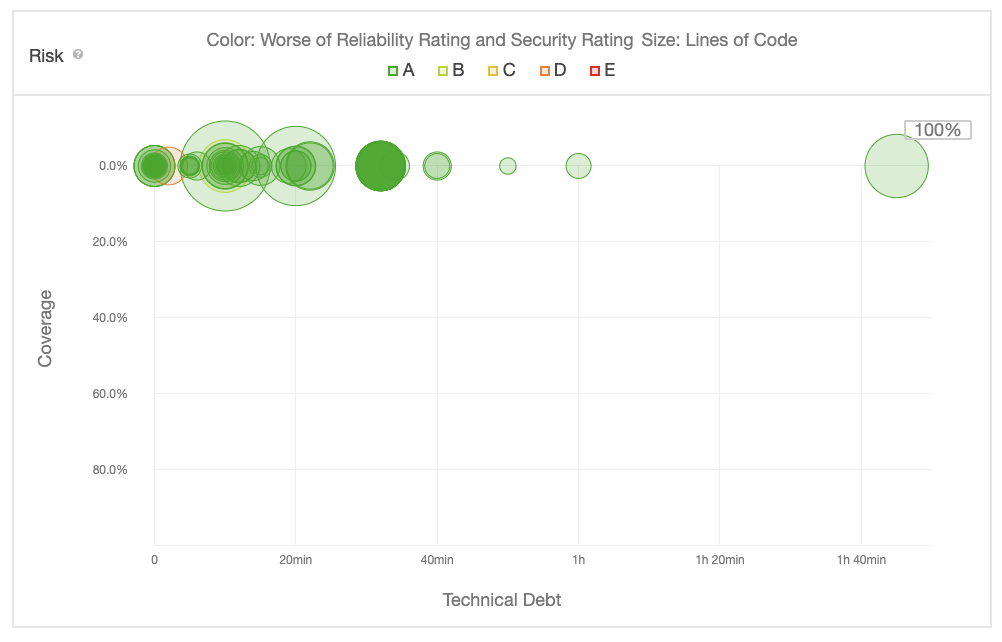}}
    \caption{Project Overview Using SonarQube.}
\end{figure}
\begin{itemize}
\item
Reliability: 
Using SonarQube, it has been observed that there are 153 bugs out of which 43 are blocker, 6 are critical, a are major and a are minor bugs. It would take
around 1 day and 7 hours (estimated time) to fix these reliability issues
(shown in Fig. 6) \cite{b4}.
\begin{figure}[hbt!]
\centering
    \frame{\includegraphics[width=\columnwidth]{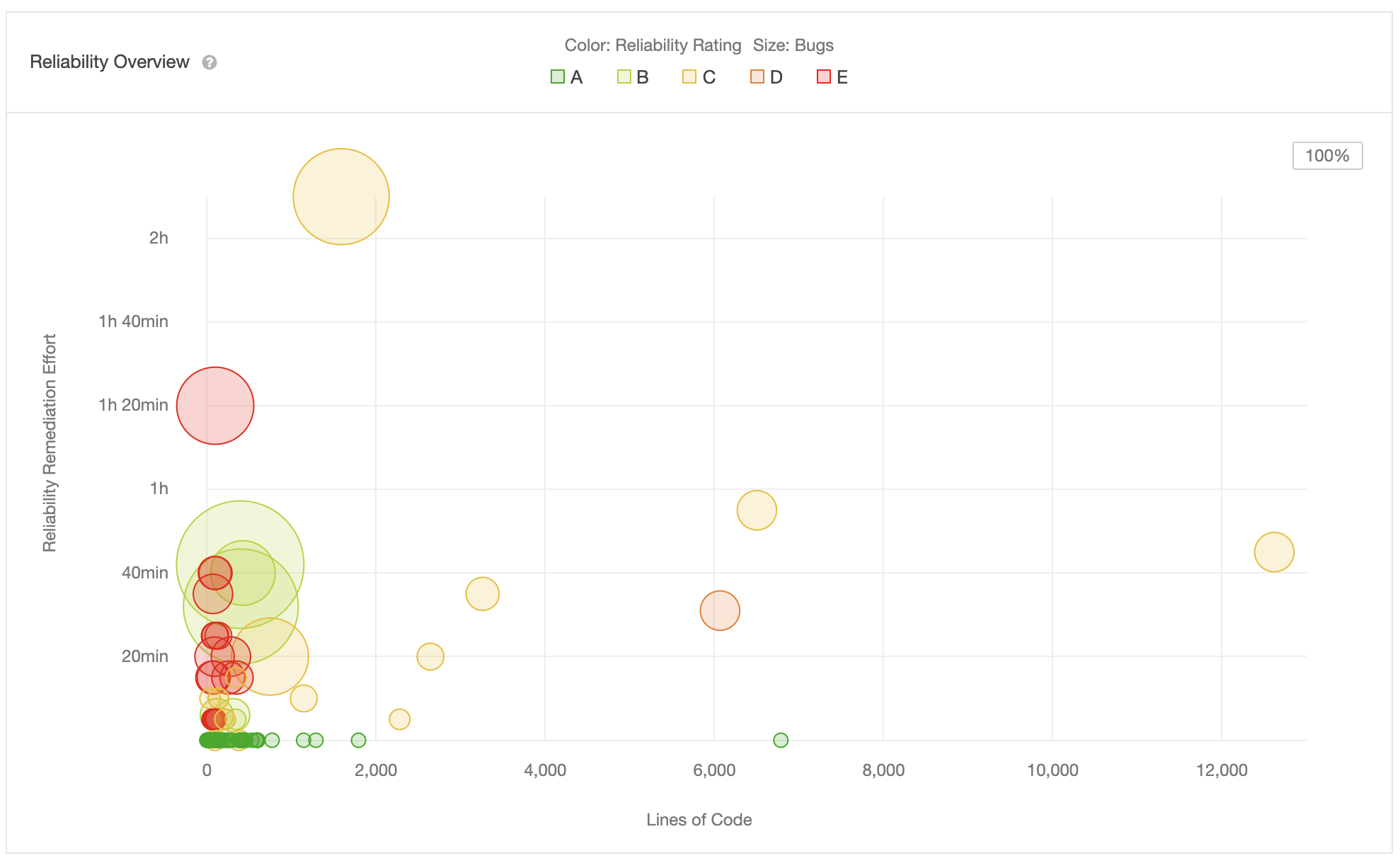}}
    \caption{Reliability Overview.} 
\end{figure}
\\
The red coloured bubbles are the worst bugs and there is 2
classes which are containing 1 of those bugs each.
\item Security: With respect to Security, there are  165 vulnerabilities
observed out of which there are 4 blocker
vulnerability which can make the whole application unstable
during production and 161 minor ones. It would take 5 days and 7 hours to fix all the vulnerability issues (depicted in Fig. 7) \cite{b4}.
\begin{figure}[hbt!]
\centering
    \frame{\includegraphics[width=\columnwidth]{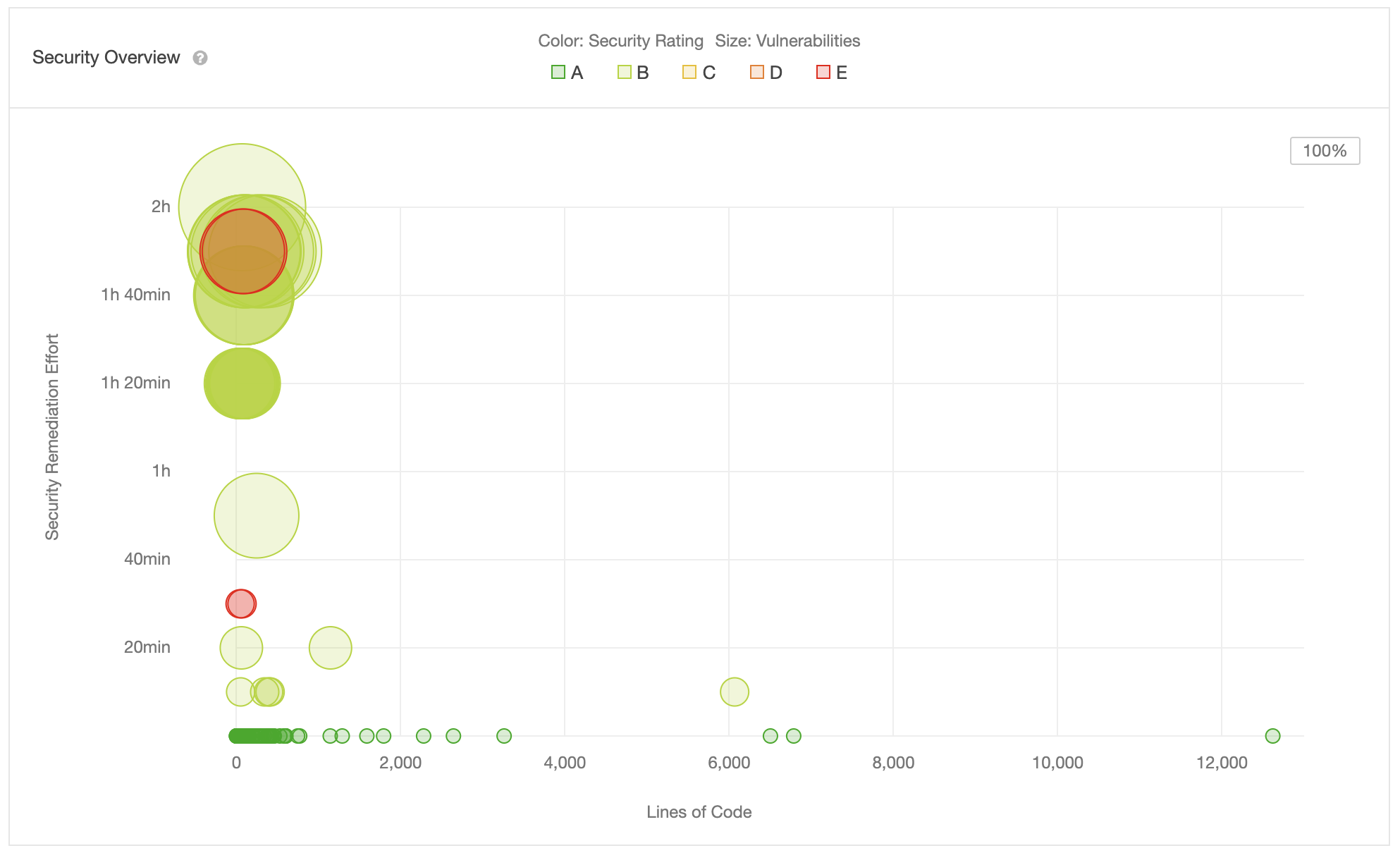}}
    \caption{Security Overview.} 
\end{figure}
\item Maintainability: 
As far as Maintainability is concerned, there are 723 smells with ratio between the cost to develop the software and
the cost to fix it (i.e., the technical debt ratio) is 0.3\% as
depicted in Fig. 8. It would take 14 days to fix all the debts \cite{b4}.
\begin{figure}[hbt!]
\centering
    \includegraphics[width=\columnwidth]{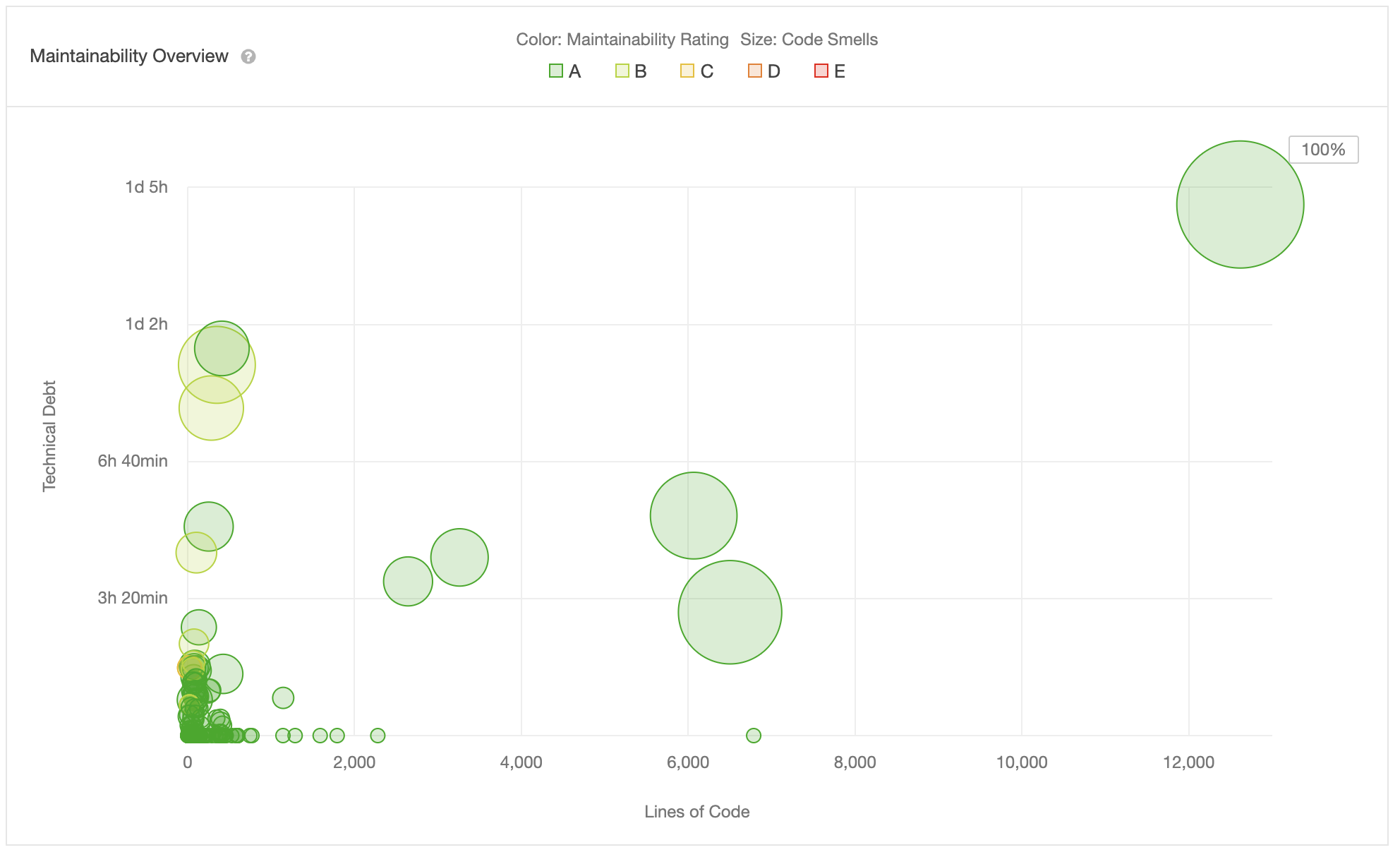}
    \caption{Maintainability Overview.} 
\end{figure}
\end{itemize}

\noindent \textbf{Quality Assessment}: SonarQube
provides a quality model which implements SQALE
methodology (Software Quality Assessment based on Life cycle
Expectations). This method mainly focuses on maintainability
issues rather than other risks involved in the project. However,
as far as our projects are concerned, we have observed that
complexity of code and maintainability was good in all. So, we
assessed quality on the basis of bugs and vulnerabilities each
possessed. Our first project contains around 2 critical bugs and 3 blocker
vulnerabilities. Whereas, second project contains 21 blocker
bugs, 153 blocker bugs and 169 blocker bugs respectively. So,
overall, ‘Booking Manager’ has been considered with the
worst Quality .
\subsection{Technical Debt Identification}
\noindent TD Identification implies detecting the occurrence of debt using tools so as to manage them accordingly. The tool in picture here is SonarQube. Our focus would be on different types of code debt detected by SonarQube. SQ helps us to identify different aspects of code debt such as Bugs, Vulnerabilities and Smells each of this is mapped to a certain quality attributes. SonarQube also helps to identify test debt and documentation debt.\\

\noindent \textbf{Project 1: Core Java 8}
\begin{itemize}
\item
Bugs: Refer figure 9; depicts 1 critical and 1 minor debt.
\begin{figure}[hbt!]
\centering
    \frame{\includegraphics[width=\columnwidth]{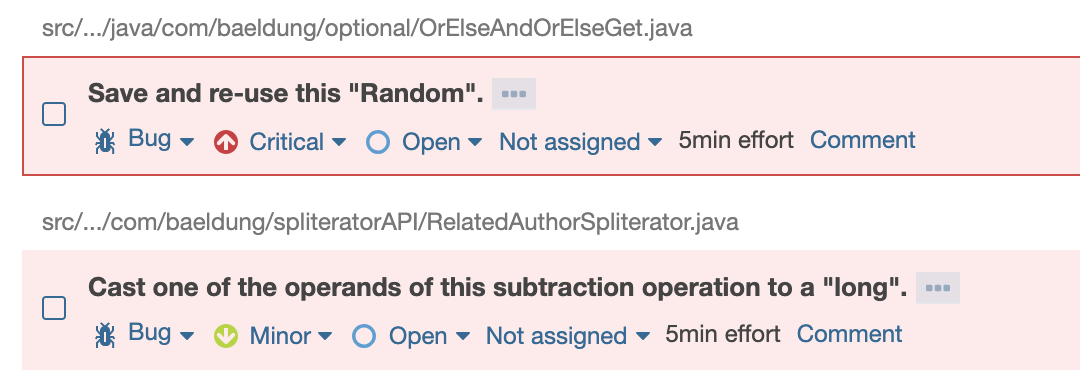}}
    \caption{Project 1 - Bugs.} 
\end{figure}
\item
Vulnerabilities: Refer figure 10; depicts 3 minor debt.
\begin{figure}[hbt!]
\centering
    \frame{\includegraphics[width=\columnwidth]{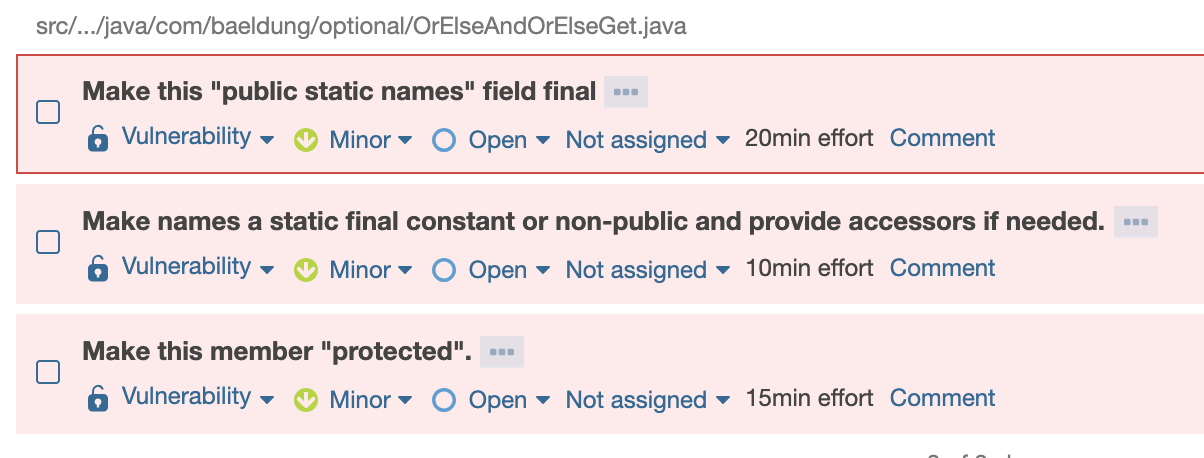}}
    \caption{Project 1 - Vulnerabilities.} 
\end{figure}
\item
Code Smells: Refer figure 11 and 12; depicts 4 critical, 2 blocker, 63 major and 1 minor debt.
\begin{figure}[hbt!]
\centering
    \frame{\includegraphics[width=\columnwidth]{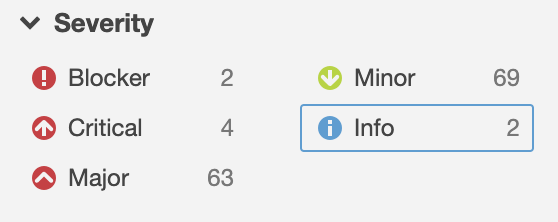}}
    \caption{Project 1 - Code Smells.} 
\end{figure}
\begin{figure}[hbt!]
\centering
    \frame{\includegraphics[width=\columnwidth]{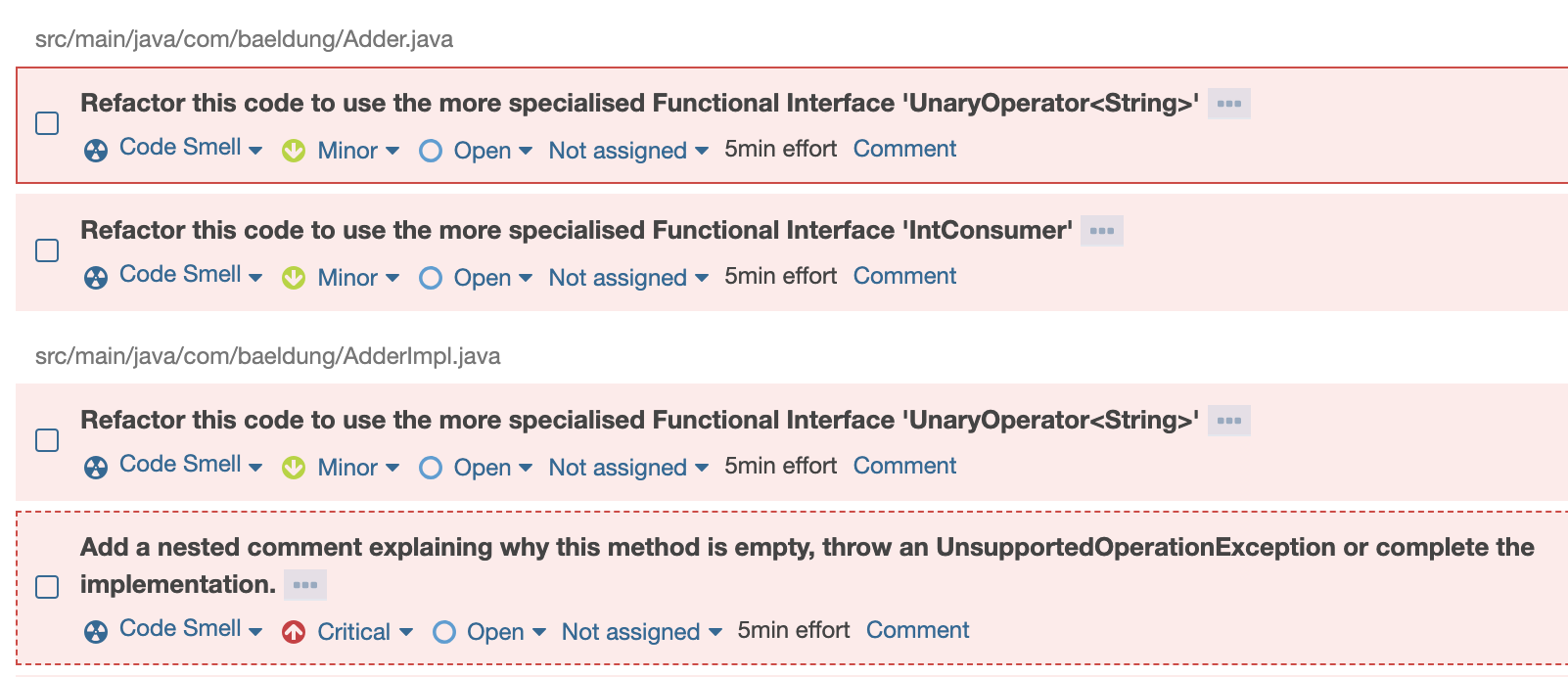}}
    \caption{Project 1 - Code Smells.} 
\end{figure}
\item
Security Hotspot: Refer figure 13; depicts 2 debt termed as security hotspot.
\begin{figure}[hbt!]
\centering
    \frame{\includegraphics[width=\columnwidth]{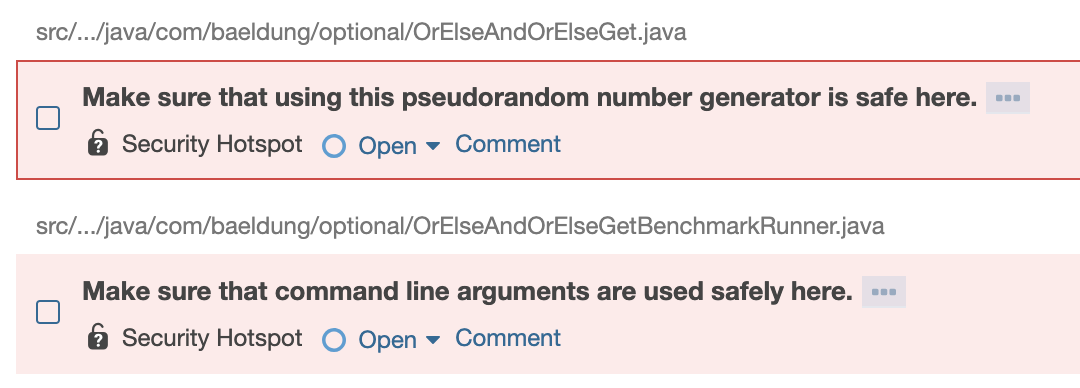}}
    \caption{Project 1 - Security Hotspot.} 
\end{figure}
\item
Duplication: Refer figure 14 and 15; depicts duplication density throughout the project.
\begin{figure}[hbt!]
\centering
    \frame{\includegraphics[width=\columnwidth]{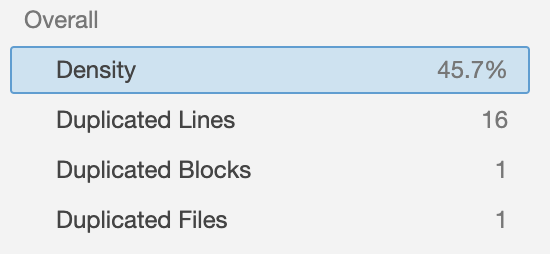}}
    \caption{Project 1 - Duplication.} 
\end{figure}
\begin{figure}[hbt!]
\centering
    \frame{\includegraphics[width=\columnwidth]{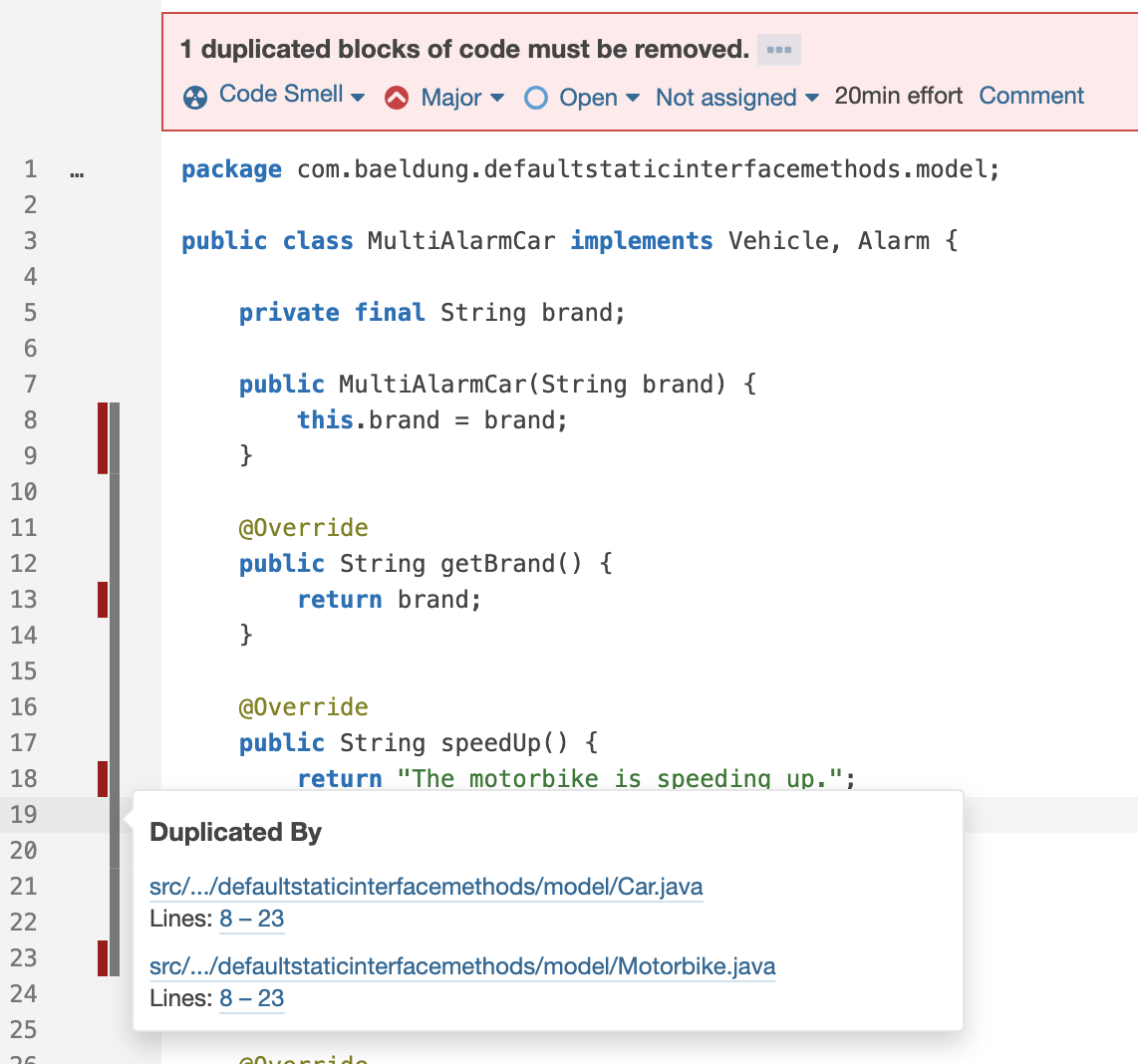}}
    \caption{Project 1 - Duplication (Sample Class).} 
\end{figure}

\item
Documentation: Refer figure 16; depicts documentation (comments) throughout the project.
\begin{figure}[hbt!]
\centering
    \frame{\includegraphics[width=\columnwidth]{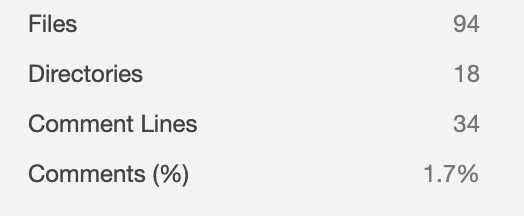}}
    \caption{Project 1 - Comments.} 
\end{figure}
\end{itemize}

\noindent \textbf{Project 2: Booking Manager}
\begin{itemize}
\item
Bugs: Refer figure 17 and 18; depicts one critical, 43 Blocker, 51 major and 1 minor debt.
\begin{figure}[hbt!]
\centering
    \frame{\includegraphics[width=\columnwidth]{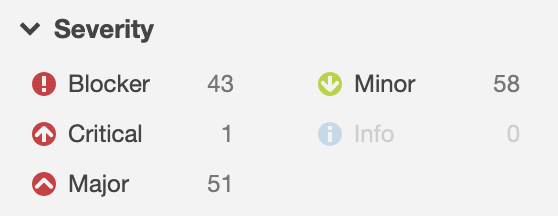}}
    \caption{Project 2 - Bugs Overview.} 
\end{figure}
\begin{figure}[hbt!]
\centering
    \frame{\includegraphics[width=\columnwidth]{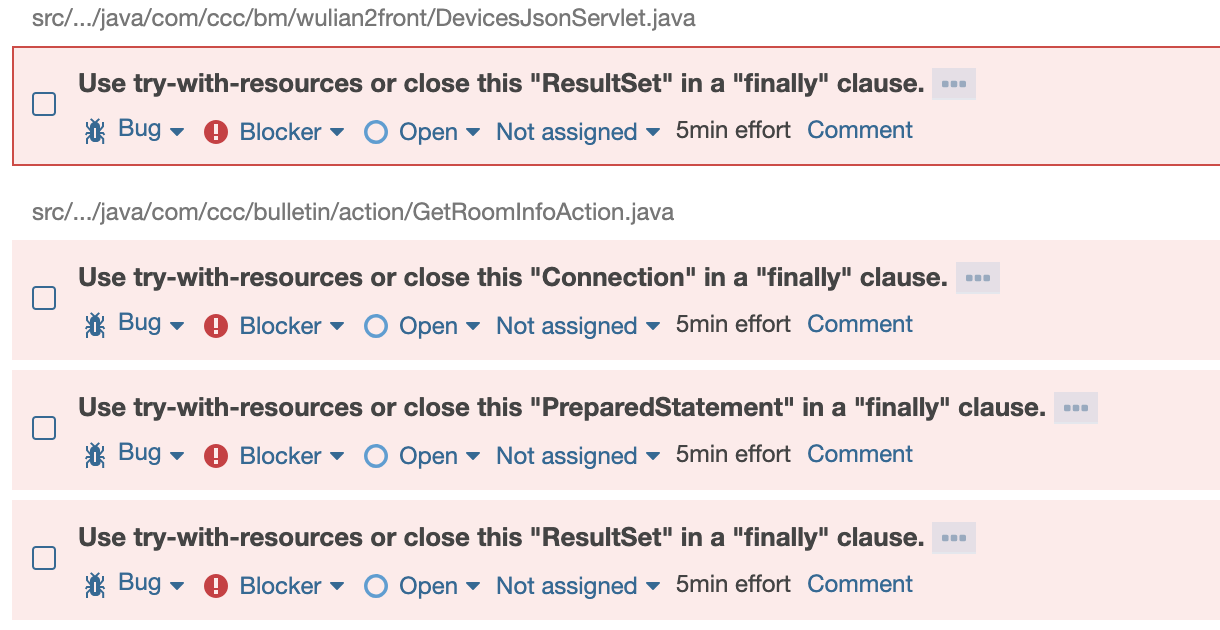}}
    \caption{Project 2 - Bugs.} 
\end{figure}
\item
Vulnerabilities: Refer figure 18 and 19; depicts 4 blocker and 161 minor debt.
\begin{figure}[hbt!]
\centering
    \frame{\includegraphics[width=\columnwidth]{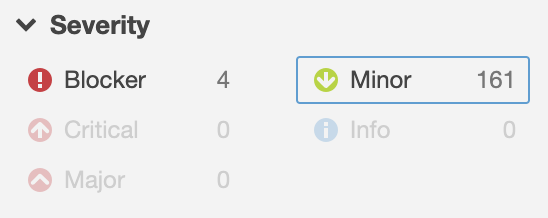}}
    \caption{Project 2 - Vulnerabilities Overview.} 
\end{figure}
\begin{figure}[hbt!]
\centering
    \frame{\includegraphics[width=\columnwidth]{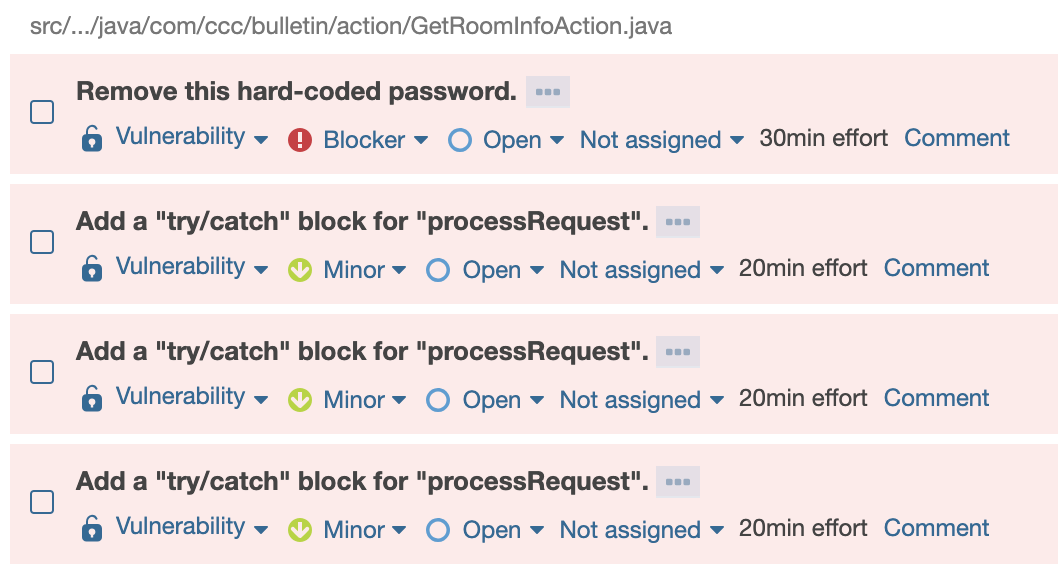}}
    \caption{Project 2 - Vulnerabilities.} 
\end{figure}
\item
Code Smells: Refer figure 20 and 21; depicts 64 critical, 6 blocker, 301 major and 331 minor debt.
\begin{figure}[hbt!]
\centering
    \frame{\includegraphics[width=\columnwidth]{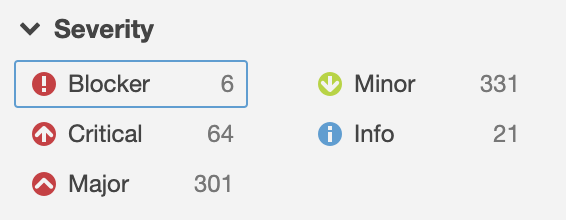}}
    \caption{Project 2 - Code Smells Overview.} 
\end{figure}
\begin{figure}[hbt!]
\centering
    \frame{\includegraphics[width=\columnwidth]{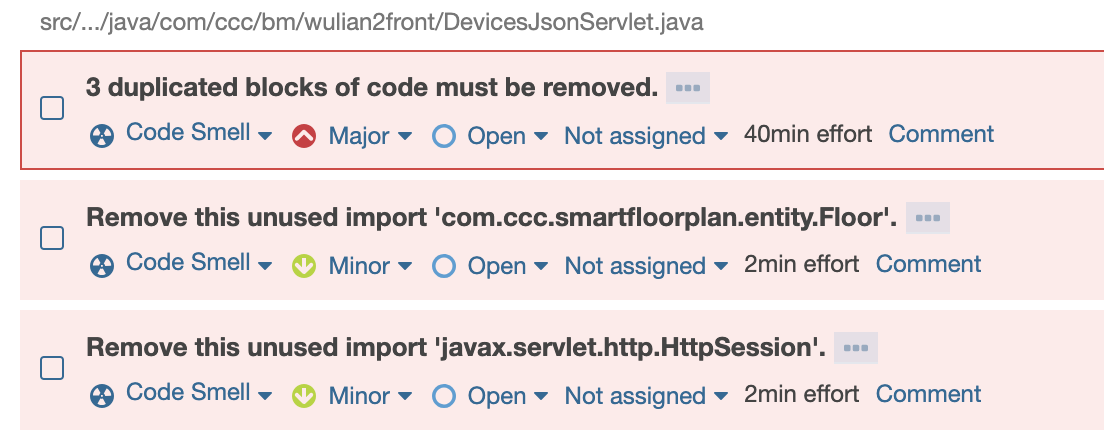}}
    \caption{Project 2 - Code Smells.} 
\end{figure}
\item
Security Hotspot: Refer figure 21 and 22; depicts 32 critical debt termed as security hotspot.

\begin{figure}[hbt!]
\centering
    \frame{\includegraphics[width=\columnwidth]{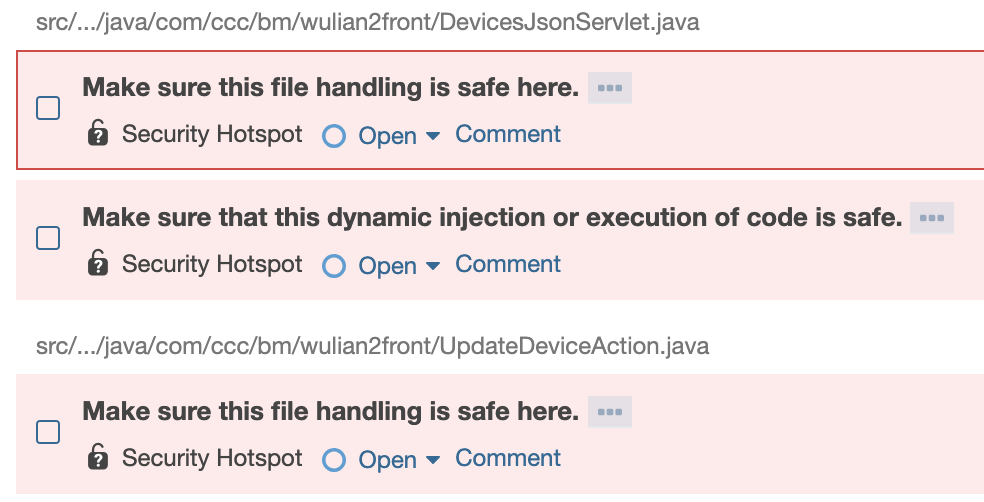}}
    \caption{Project 2 - Security Hotspot.} 
\end{figure}
\item
Duplication: Refer figure 23 and 24; depicts duplication density throughout the project.
\begin{figure}[hbt!]
\centering
    \frame{\includegraphics[width=\columnwidth]{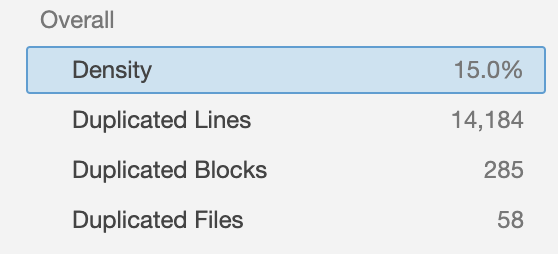}}
    \caption{Project 2 - Duplication.} 
\end{figure}
\begin{figure}[hbt!]
\centering
    \frame{\includegraphics[width=\columnwidth]{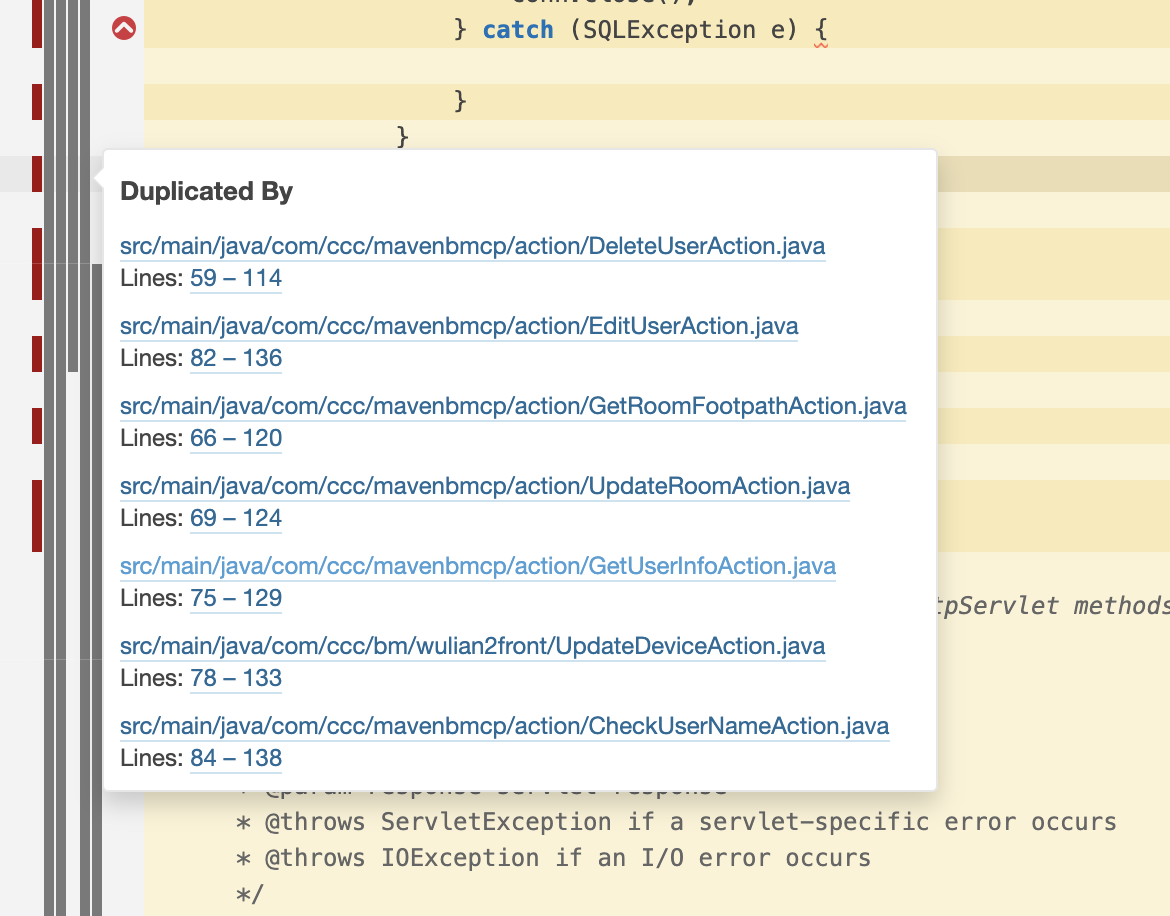}}
    \caption{Project 2 - Duplication (Sample Class).} 
\end{figure}
\item
Documentation: Refer figure 25; depicts documentation (comments) throughout the project.
\begin{figure}[hbt!]
\centering
    \frame{\includegraphics[width=\columnwidth]{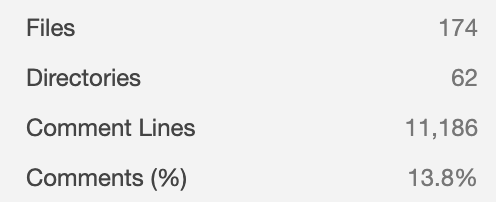}}
    \caption{Project 2 - Comments.} 
\end{figure}
\end{itemize}
\noindent \textbf{Mapping of TD Items and Dimensions}: TD items identified above have been mapped to their
respective dimension; refer table III.
\begin{center}
\begin{table}[hbt!]
\centering
\caption{Mapping of TD Items and Dimensions} % title name of the table
\begin{tabular}{ |c|c|c|c| } 
\hline
Project & TD Items & TD Dimension \\
\hline
\multirow{6}{4em}{Core Java 8} & Bugs & Code Debt \\ 
& Vulnerabilities & Code Debt \\ 
& Code Smells & Code Debt \\ 
& Security Hotspot & Code Debt \\ 
& JUnit Test coverage & Test Debt \\ 
& Comments Completeness & Documentation Debt \\
\hline
\multirow{6}{4em}{Booking Manager} & Bugs & Code Debt \\ 
& Vulnerabilities & Code Debt \\ 
& Code Smells & Code Debt \\ 
& Security Hotspot & Code Debt \\ 
& JUnit Test coverage & Test Debt \\ 
& Comments Completeness & Documentation Debt \\
\hline
\end{tabular}
\label{tab:PPer}
\end{table}
\end{center}
\subsection{Technical Debt Representation}

\noindent \textbf{Project 1: Core Java 8}
Refer table 4, 5 and 6 for TD Representation.\\
\noindent \textbf{Project 2: Booking Manager}
Refer table 7, 8, 9, 10 and 11 for TD Representation.\\
\begin{center}
\begin{table}[hbt!]
\centering
\caption{TD Representation}
\begin{tabular}{ |c|c|c|c| } 
\hline
ID & 1.1  \\
\hline
Name & Bug \\ 
\hline
Location & src/.../optional/OrElseAndOrElseGet.java

 \\ 
\hline
Responsible/author & Not Assigned \\ 
\hline
Dimension & Code Debt \\ 
\hline
Date/time & Apr 14, 2019 15 : 43 :18 \\ 
\hline
Context & "Random" objects should be reused
 \\ 
\hline
Propagation rule &  May produce non accepted results; JDK dependent\\ 
\hline
Intentionality & N/A \\ 
\hline
\end{tabular}
\label{tab:PPer}
\end{table}
\end{center}

\begin{center}
\begin{table}[hbt!]
\centering
\caption{TD Representation}
\begin{tabular}{ |c|c|c|c| } 
\hline
ID & 1.2  \\
\hline
Name & Code Smell \\ 
\hline
Location & src/../AdderImpl.java

 \\ 
\hline
Responsible/author & Not Assigned \\ 
\hline
Dimension & Code Debt \\ 
\hline
Date/time & Apr 14, 2019 15 : 23 :18 \\ 
\hline
Context & Methods should not be empty
 \\ 
\hline
Propagation rule & Can cause unexpected behavior in production. \\ 
\hline
Intentionality & No \\ 
\hline
\end{tabular}
\label{tab:PPer}
\end{table}
\end{center}
\begin{center}
\begin{table}[hbt!]
\centering
\caption{TD Representation}
\begin{tabular}{ |c|c|c|c| } 
\hline
ID & 1.3  \\
\hline
Name & Code Smell \\ 
\hline
Location & src/.../application/Application.java
 \\ 
\hline
Responsible/author & Not Assigned \\ 
\hline
Dimension & Code Debt \\ 
\hline
Date/time & Apr 14, 2019 15 : 37 :18 \\ 
\hline
Context &  Logging\\ 
\hline
Propagation rule & Useful for debugging \\ 
\hline
Intentionality & N/A \\ 
\hline
\end{tabular}
\label{tab:PPer}
\end{table}
\end{center}
\begin{center}
\begin{table}[hbt!]
\centering
\caption{TD Representation}
\begin{tabular}{ |c|c|c|c| } 
\hline
ID & 2.1  \\
\hline
Name & Bug \\ 
\hline
Location & src/.../wulian2front/DevicesJsonServlet.java
 \\ 
\hline
Responsible/author & Not Assigned \\ 
\hline
Dimension & Code Debt \\ 
\hline
Date/time & Apr 14, 2019 15 : 43 :18 \\ 
\hline
Context & Failure to properly close resources \\ 
\hline
Propagation rule &  This can lead to denial of service\\ 
\hline
Intentionality & No \\ 
\hline
\end{tabular}
\label{tab:PPer}
\end{table}
\end{center}
\begin{center}
\begin{table}[hbt!]
\centering
\caption{TD Representation}
\begin{tabular}{ |c|c|c|c| } 
\hline
ID & 2.2  \\
\hline
Name & Bug \\ 
\hline
Location & src/.../DataTables-1.10.6/js/jquery.dataTables.js
 \\ 
\hline
Responsible/author & Not Assigned \\ 
\hline
Dimension & Code Debt \\ 
\hline
Date/time & Apr 14, 2019 15 : 23 :18 \\ 
\hline
Context & Mixing up the order of operations \\ 
\hline
Propagation rule & May hamper the overall behaviour \\ 
\hline
Intentionality & No \\ 
\hline
\end{tabular}
\label{tab:PPer}
\end{table}
\end{center}
\begin{center}
\begin{table}[hbt!]
\centering
\caption{TD Representation}
\begin{tabular}{ |c|c|c|c| } 
\hline
ID & 2.3  \\
\hline
Name & Bug \\ 
\hline
Location & src/.../action/GetRoomInfoAction.java
 \\ 
\hline
Responsible/author & Not Assigned \\ 
\hline
Dimension & Code Debt \\ 
\hline
Date/time & Apr 14, 2019 15 : 37 :18 \\ 
\hline
Context &  Non-serializable objects\\ 
\hline
Propagation rule & Objects in the session can throw error \\ 
\hline
Intentionality & No \\ 
\hline
\end{tabular}
\label{tab:PPer}
\end{table}
\end{center}
\begin{center}
\begin{table}[hbt!]
\centering
\caption{TD Representation}
\begin{tabular}{ |c|c|c|c| } 
\hline
ID & 2.4  \\
\hline
Name & Security Hotsapot \\ 
\hline
Location & src/.../wulian2front/DevicesJsonServlet.java
 \\ 
\hline
Responsible/author & Not Assigned \\ 
\hline
Dimension & Code Debt \\ 
\hline
Date/time & Apr 15, 2019 15 : 43 :18 \\ 
\hline
Context & File Handling \\ 
\hline
Propagation rule & Exposing a file's content is dangerous\\ 
\hline
Intentionality & No \\ 
\hline
\end{tabular}
\label{tab:PPer}
\end{table}
\end{center}
\begin{center}
\begin{table}[hbt!]
\centering
\caption{TD Representation}
\begin{tabular}{ |c|c|c|c| } 
\hline
ID & 2.5  \\
\hline
Name & Security Hotspot \\ 
\hline
Location & src/.../ccc/bm/wulian2front/DevicesJsonServlet.java
 \\ 
\hline
Responsible/author & Not Assigned \\ 
\hline
Dimension & Code Debt \\ 
\hline
Date/time & Apr 14, 2019 15 : 23 :18 \\ 
\hline
Context &  Dynamic Code Execution\\ 
\hline
Propagation rule & Dangerous to execute unknown code \\ 
\hline
Intentionality & No \\ 
\hline
\end{tabular}
\label{tab:PPer}
\end{table}
\end{center}
\noindent Please Note: For some TD items, Intentionality is N/A since the
intentional debts observed are of documentation debt only. Also,
for the propagation rules its only the blocker code smells that
mainly affects the whole project so for the rest of the TD items
it is assumed to be N/A. This has been followed in Appendix as
well.

\subsection{Technical debt estimation}
Technical debt estimation basically implies the effort required to fix the identified debts. We here, use SonarQube to estimate the efforts required for each of the different TD items \cite{b5}\cite{b6}.

\noindent \textbf{Project 1: Core Java 8}
\noindent Refer figure 28, 29 and 30 for the estimates.\\
\begin{figure}[hbt!]
\centering
    \frame{\includegraphics[width=\columnwidth]{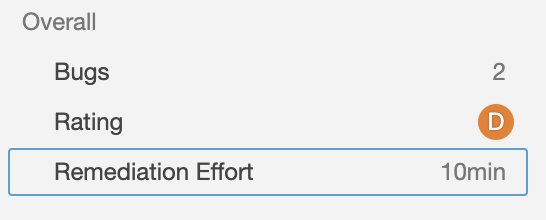}}
    \caption{Project 1 - Estimates.} 
\end{figure}
\begin{figure}[hbt!]
\centering
    \frame{\includegraphics[width=\columnwidth]{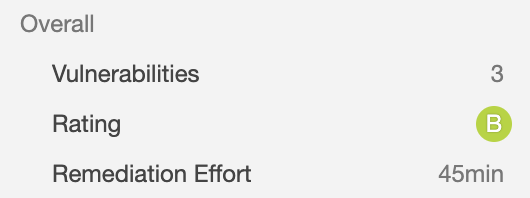}}
    \caption{Project 1 - Estimates.} 
\end{figure}
\begin{figure}[hbt!]
\centering
    \frame{\includegraphics[width=\columnwidth]{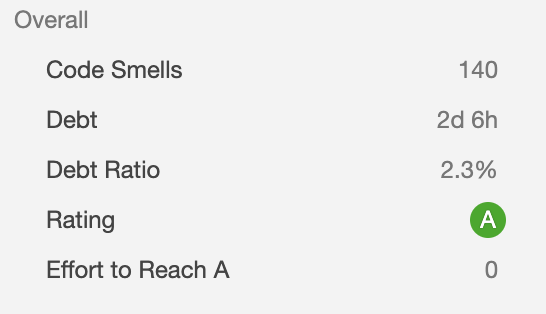}}
    \caption{Project 1 - Estimates.} 
\end{figure}
\noindent \textbf{Project 2: Booking Manager}
\noindent Refer figure 31, 32 and 33 for the estimates.
\begin{figure}[hbt!]
\centering
    \frame{\includegraphics[width=\columnwidth]{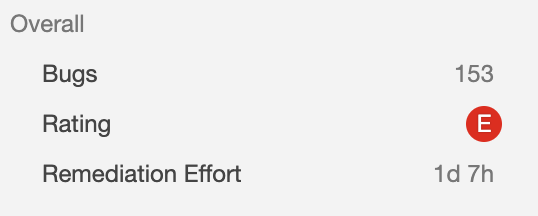}}
    \caption{Project 2 - Estimates.} 
\end{figure}
\begin{figure}[hbt!]
\centering
    \frame{\includegraphics[width=\columnwidth]{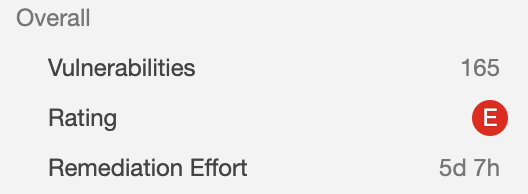}}
    \caption{Project 2 - Estimates.} 
\end{figure}
\begin{figure}[hbt!]
\centering
    \frame{\includegraphics[width=\columnwidth]{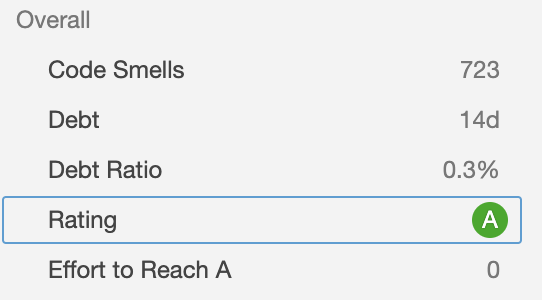}}
    \caption{Project 2 - Estimates.} 
\end{figure}

\subsection{Technical Debt Monitoring}
\noindent It is one of the Technical Debt management activity which
controls the changes in the cost and benefit of the remaining
debt items as the time passes by. There are various approaches
to monitor TD:\\
\noindent \textbf{Threshold-based approach} : specify thresholds for TD related
quality metrics, and issue warnings if these thresholds are
exceeded. \\
\noindent \textbf{TD propagation tracking} : Track the impact of TD via
dependencies between the parts of the system where TD have
been identified and other parts of a system.\\
\noindent \textbf{Planned check} : constantly measure identified TD and track the
TD changes.\\
\noindent \textbf{TD monitoring with quality attribute focus} : Monitor the
change of quality attributes occurring at the expense of TD (e.g.,
stability).\\
\noindent \textbf{TD plot} : Plot various aggregated TD measures over time and
investigate the TD trends based on the shape of the plot curve.

Here, we use SonarQube to monitor Technical Debt using TD Plot approach. SonarQube provides dashboard with different aspects and there respective estimates.
\\
\begin{figure}[hbt!]
\centering
    \frame{\includegraphics[width=\columnwidth]{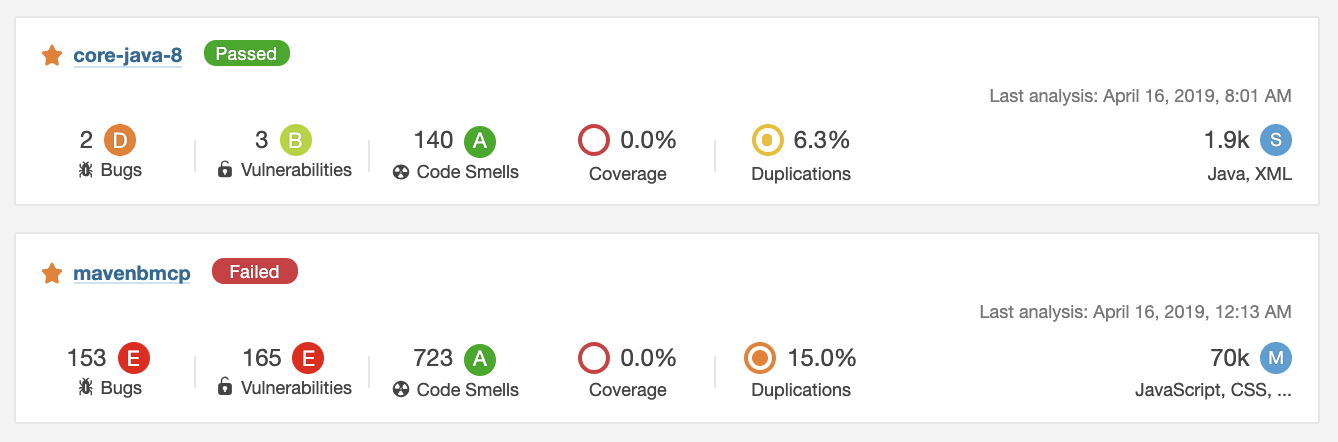}}
    \caption{SonarQube Dashboard.} 
\end{figure}
\noindent \textbf{Project 1: Core Java 8} - 
Refer figure 34 which depicts the dashboard for Project 1.
\begin{figure}[hbt!]
\centering
    \frame{\includegraphics[width=\columnwidth]{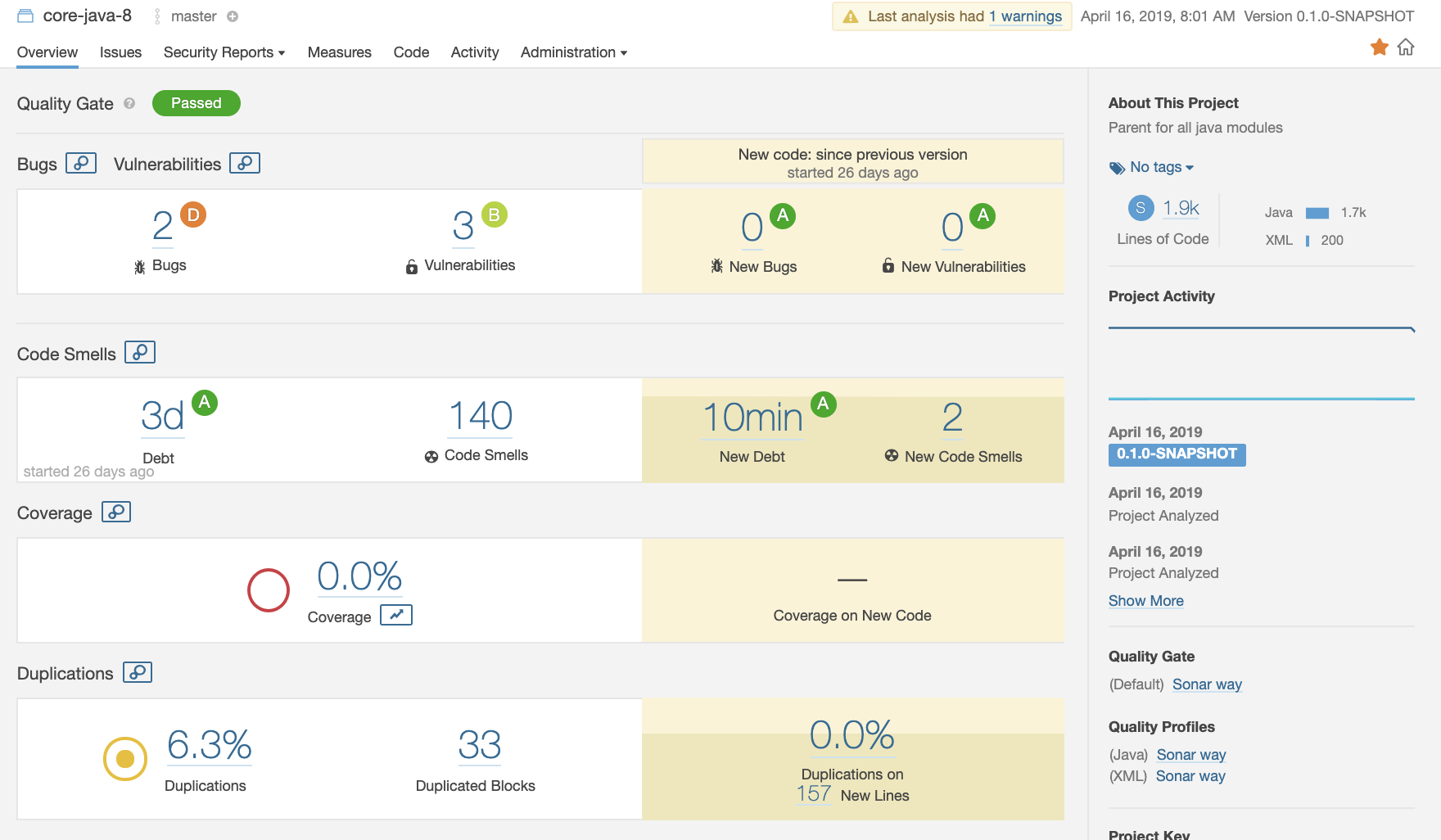}}
    \caption{Project 1 - Dashboard.} 
\end{figure}

\noindent \textbf{Project 2: Booking Manager} - 
Refer figure 35 which depicts the dashboard for Project 2.

\begin{figure}[hbt!]
\centering
    \frame{\includegraphics[width=\columnwidth]{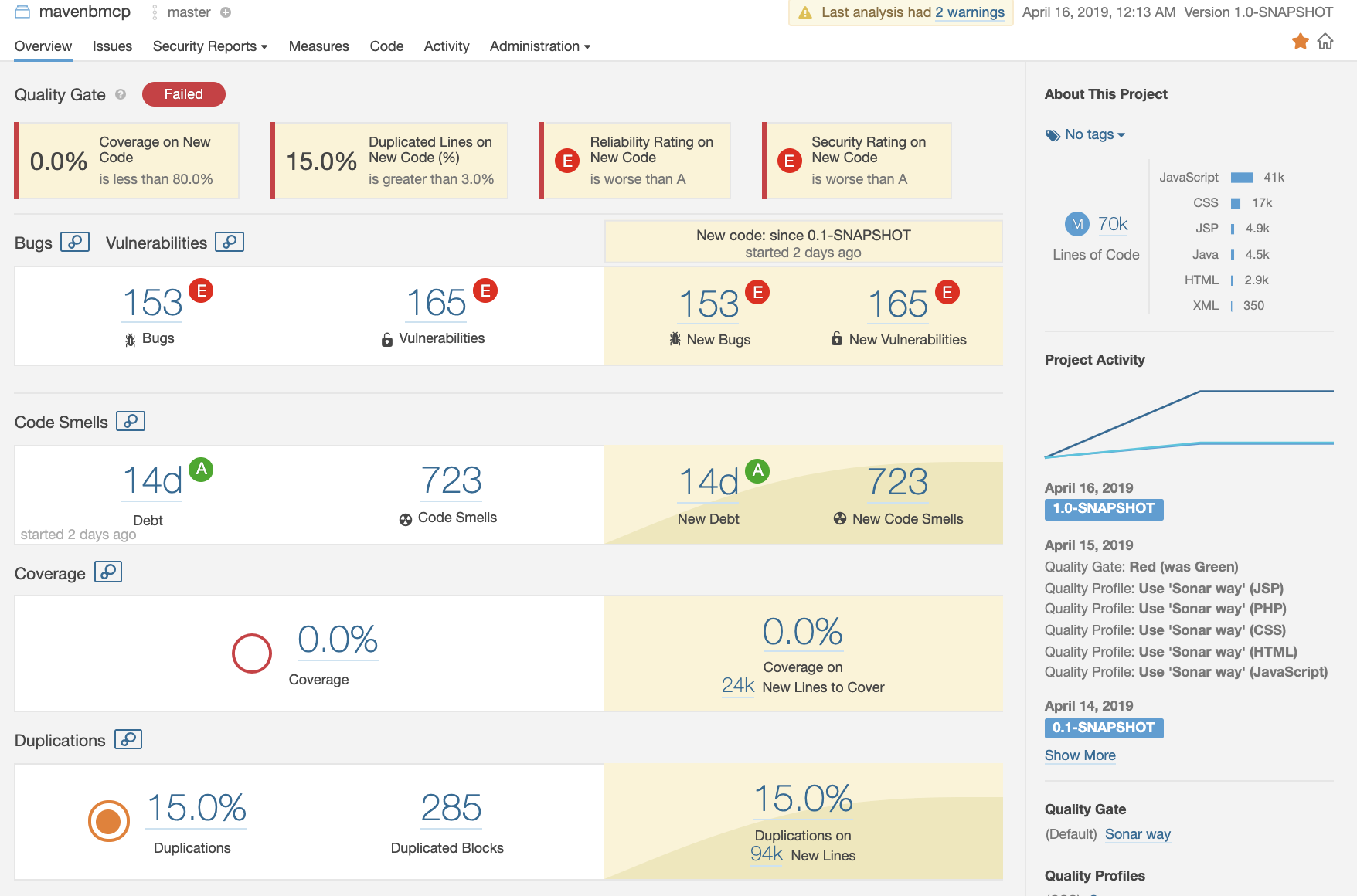}}
    \caption{Project 2 - Dashboard.} 
\end{figure}

\noindent The dashboard shows all the different TD dimensions with rating. Rating of A implies good quality code. Upon clicking on each of the dimensions, the dashboard gives detailed description of TD along with the efforts estimates and resolution techniques.
\\
Here we can say that the Project 2 has bad overall quality rating compared to project 1.

\subsection{Technical debt repayment}

\noindent Technical debt repayment is one of significant TD Management
techniques because paying back the principal will keep technical
debt under control. It also allows the programmer to focus on
other issues such as developing the software or adding new
features. In addition, it will prevent TD from being accumulated
and keep paying the interest for a long time \cite{b9}. There are several techniques to repay TD such as Refactoring, Rewriting and Automation.
After analyzing TD identified in each project, in this section, we
proposed aforementioned techniques to repay the debt occurred
in each project.
\\
\noindent \textbf{Project 1: Core Java}: Figure 36, 37, 38, 39, 40 and 41 
shows TD item and  refactoring techniques suggested by SonarQube.\\

\begin{figure}[hbt!]
\centering
    \frame{\includegraphics[width=\columnwidth]{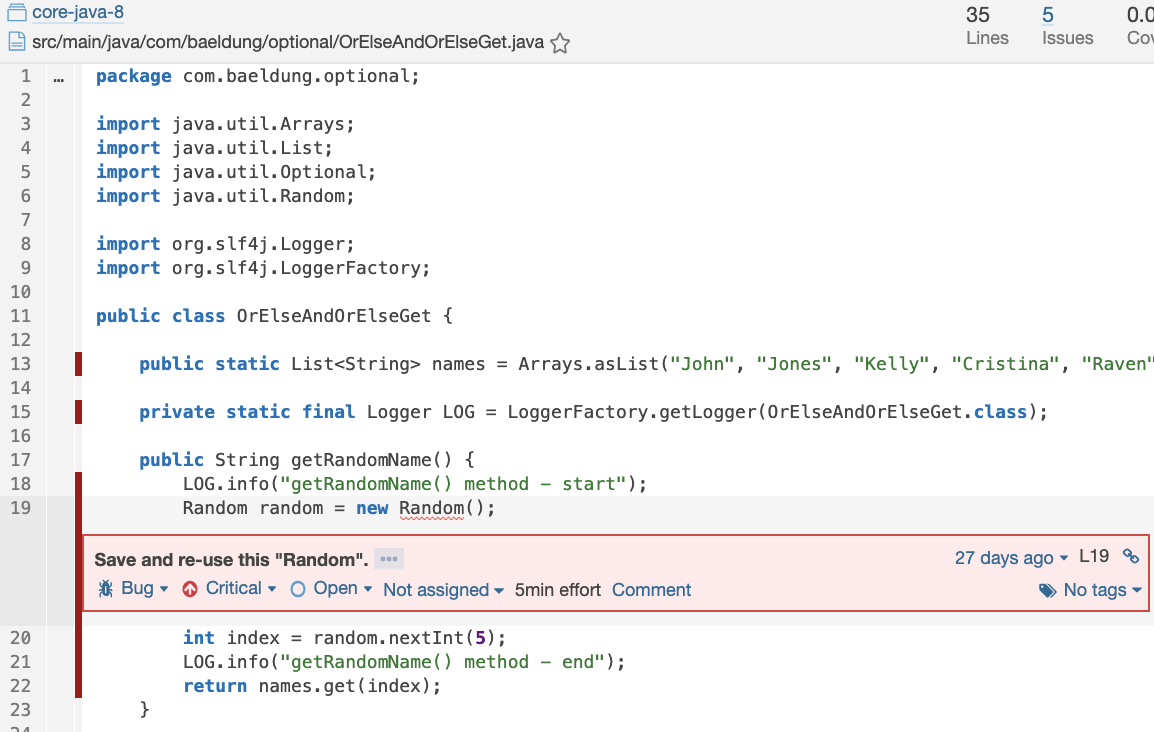}}
    \caption{TD Item.} 
\end{figure}
\begin{figure}[hbt!]
\centering
    \frame{\includegraphics[width=\columnwidth]{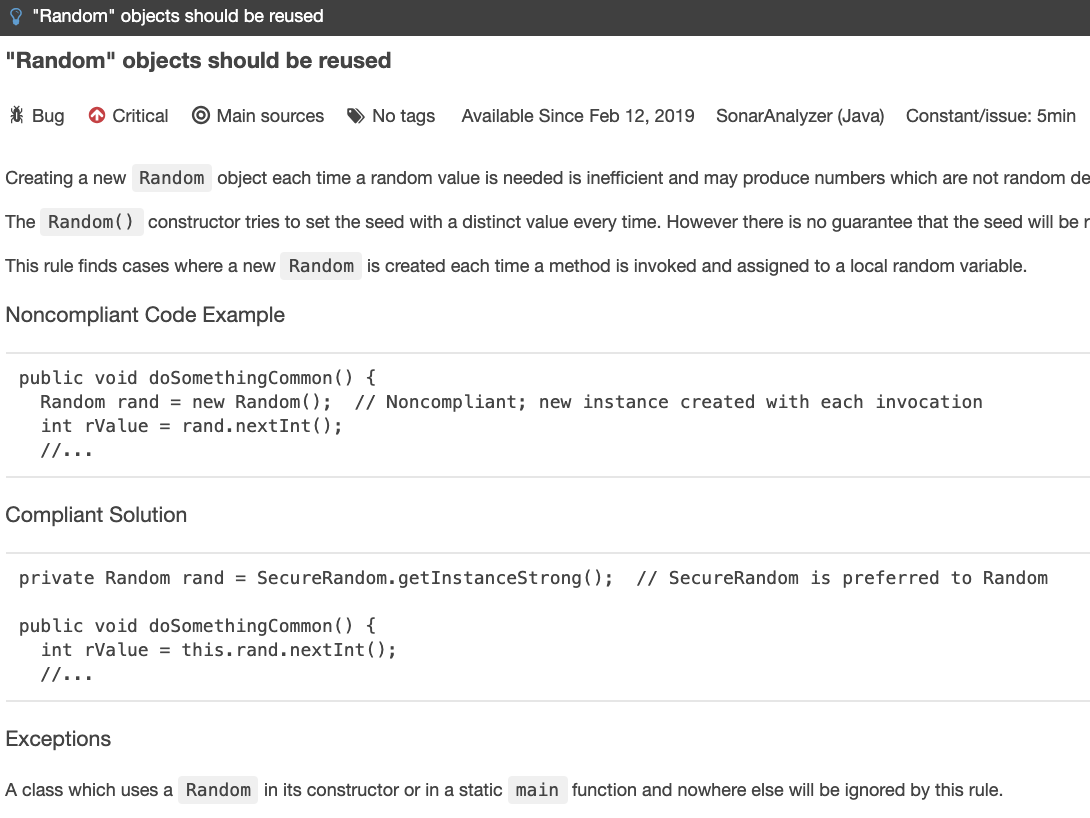}}
    \caption{Refactoring suggestion.} 
\end{figure}

\begin{figure}[hbt!]
\centering
    \frame{\includegraphics[width=\columnwidth]{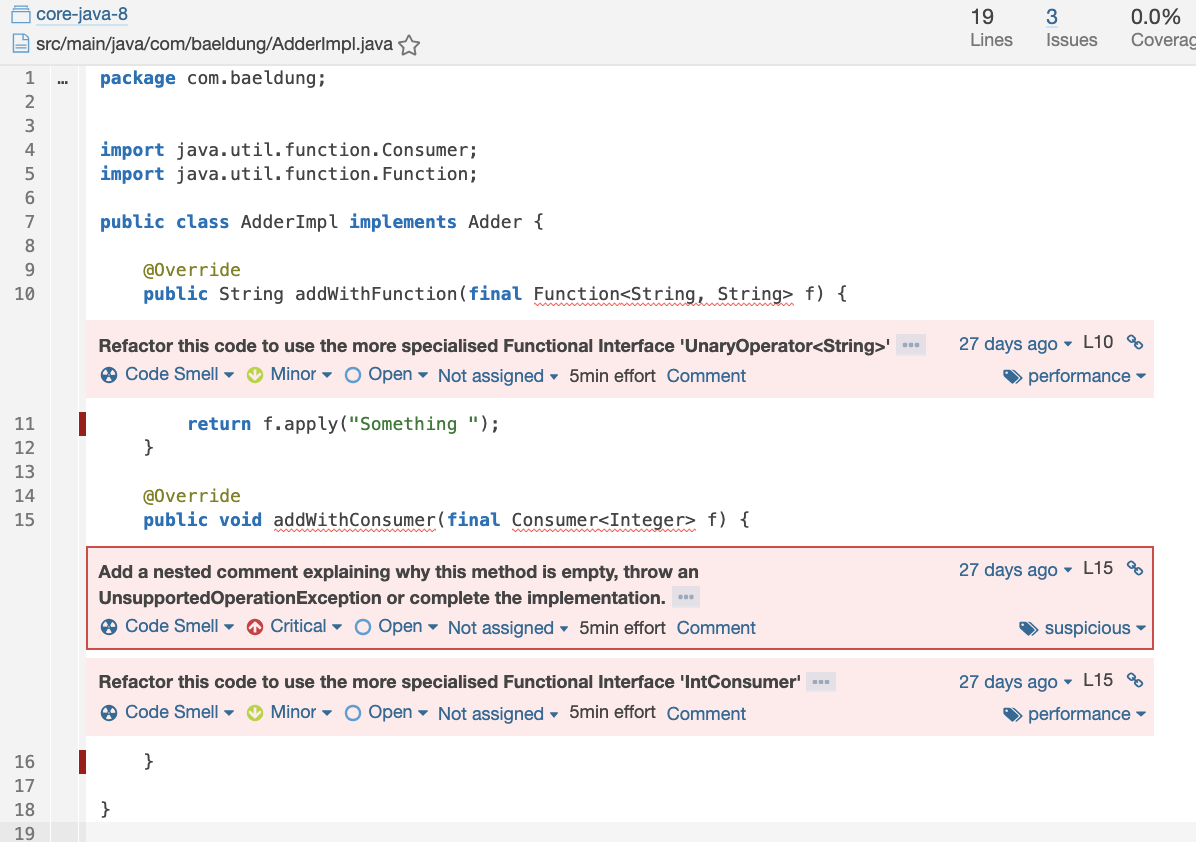}}
    \caption{TD Item.} 
\end{figure}
\begin{figure}[hbt!]
\centering
    \frame{\includegraphics[width=\columnwidth]{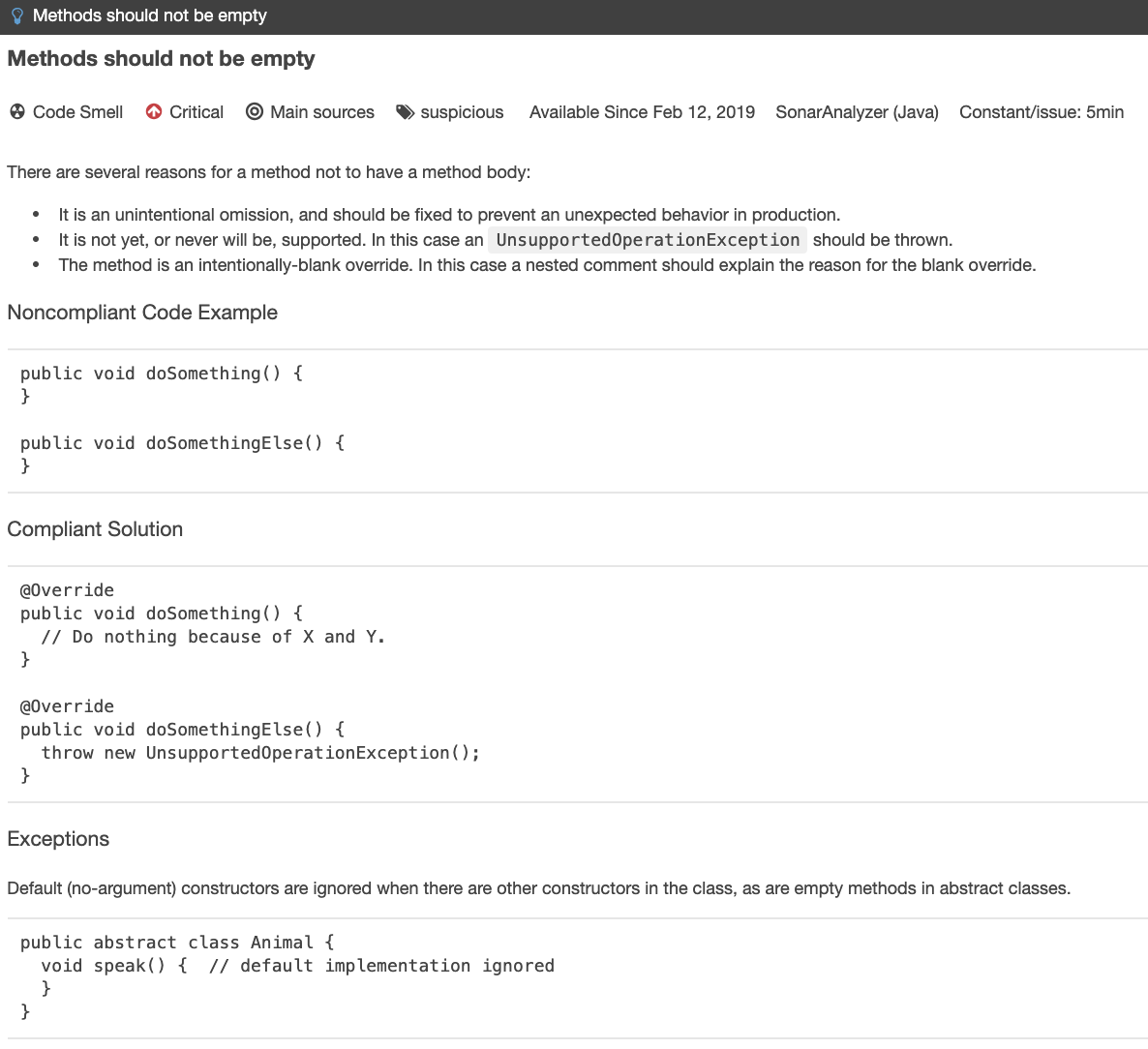}}
    \caption{Refactoring suggestion.} 
\end{figure}

\begin{figure}[hbt!]
\centering
    \frame{\includegraphics[width=\columnwidth]{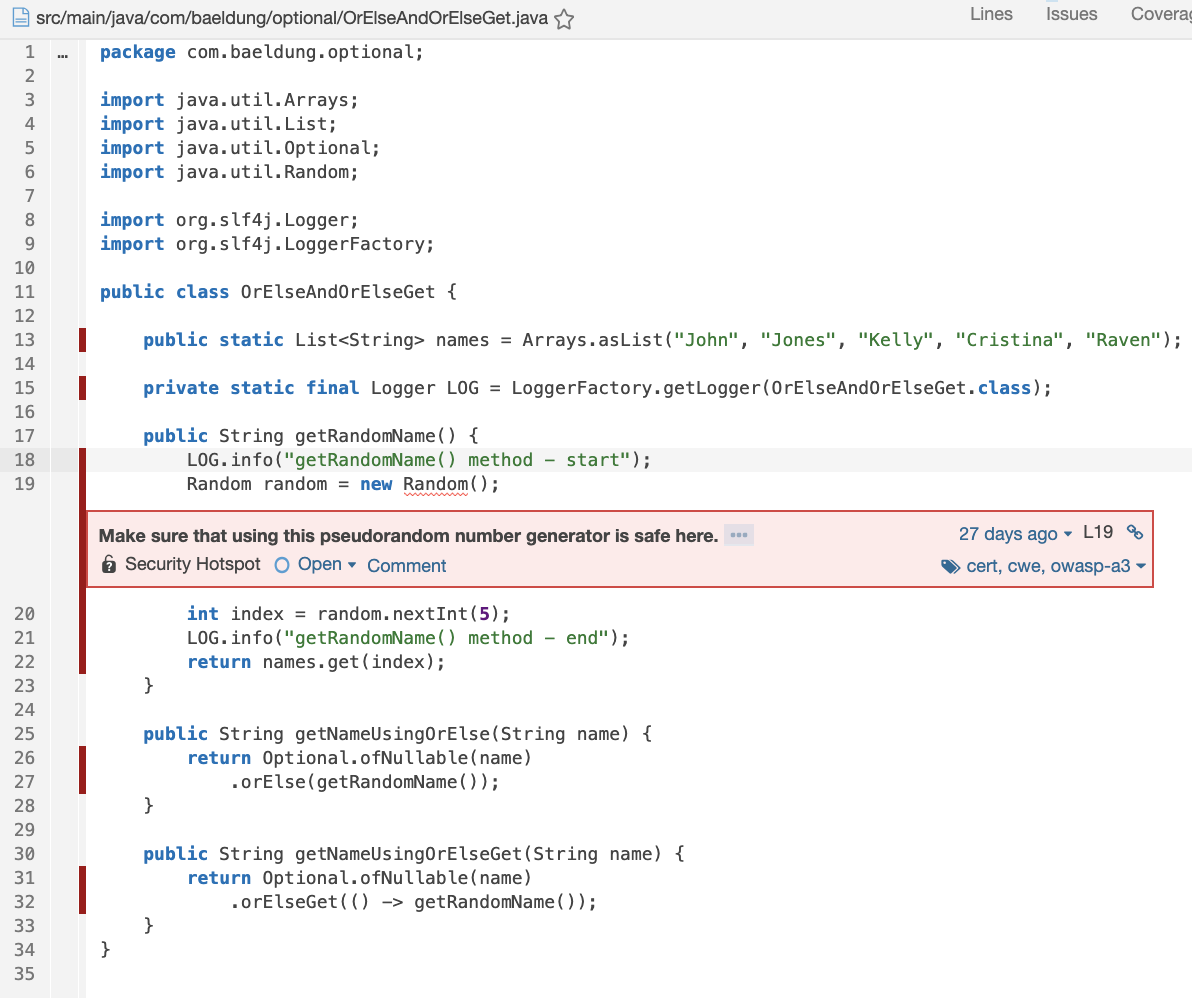}}
    \caption{TD Item.} 
\end{figure}
\begin{figure}[hbt!]
\centering
    \frame{\includegraphics[width=\columnwidth]{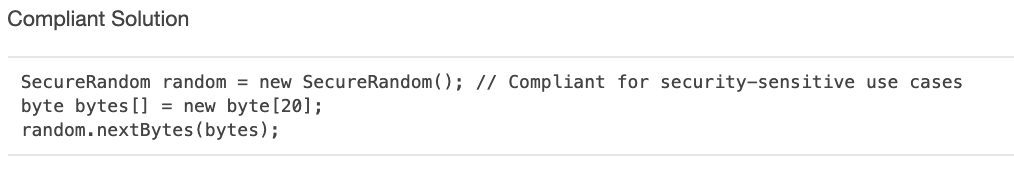}}
    \caption{Refactoring suggestion.} 
\end{figure}

\noindent \textbf{Project 2: Booking Manager}: Figure 42, 43, 44 and 45 
shows TD item and  refactoring techniques suggested by SonarQube.\\
\begin{figure}[hbt!]
\centering
    \frame{\includegraphics[width=\columnwidth]{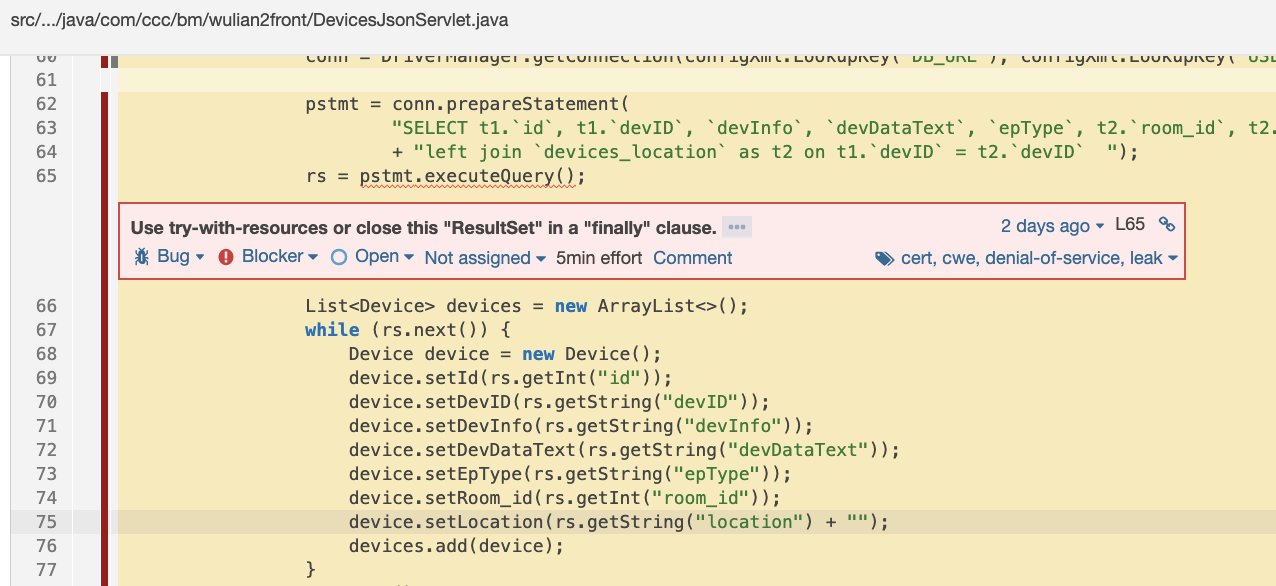}}
    \caption{TD Item.} 
\end{figure}
\begin{figure}[hbt!]
\centering
    \frame{\includegraphics[width=\columnwidth]{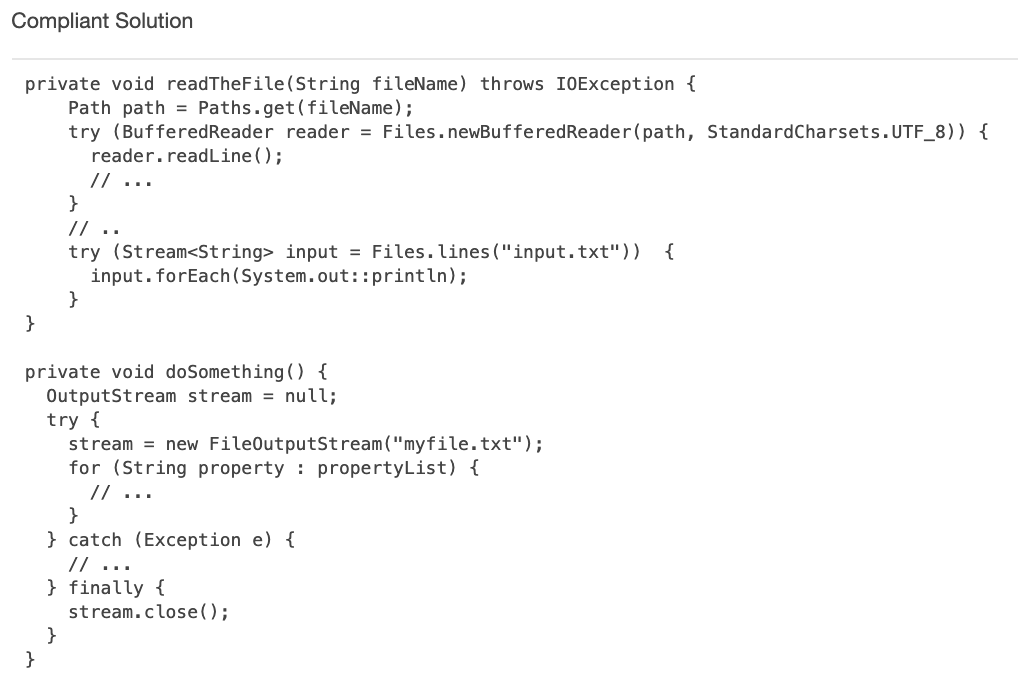}}
    \caption{Refactoring suggestion.} 
\end{figure}
\begin{figure}[hbt!]
\centering
    \frame{\includegraphics[width=\columnwidth]{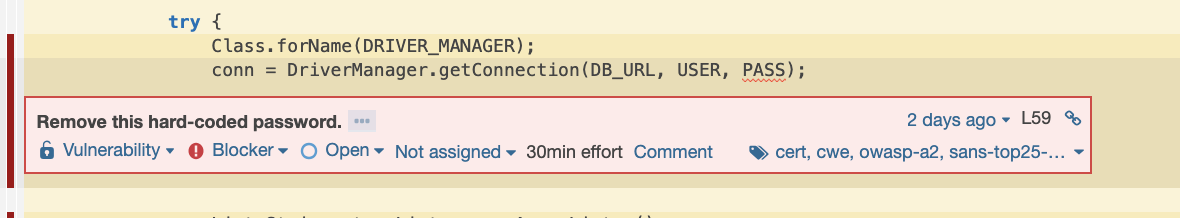}}
    \caption{TD Item.} 
\end{figure}
\begin{figure}[hbt!]
\centering
    \frame{\includegraphics[width=\columnwidth]{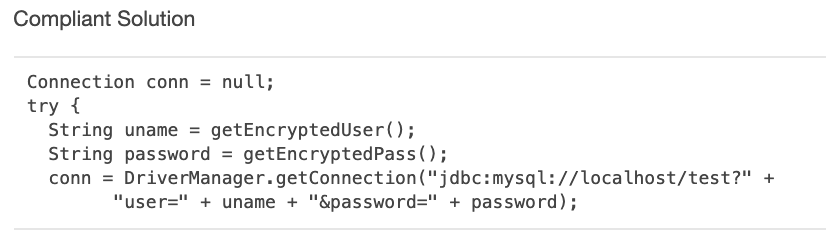}}
    \caption{Refactoring suggestion.} 
\end{figure}

\subsection{Technical debt prevention} 
\noindent Technical Debt prevention is one of the Technical Debt
Management activity that prevents potential TD from being
incurred. However, there is no such tool for TD prevention
because it is mainly supported by software development process
improvement. Nevertheless, there is a tool named Umple, which
helps to prevent TD by supporting model-oriented
programming.
There are four different approaches to prevent TD potentially: \\
\noindent \textbf{Development process improvement}: improve current
development processes to prevent the occurrences of certain
types of TD. Development process can notably be improved by
adopting continuous integration in the software development
process. \\
\noindent \textbf{Architecture decision-making support}: evaluate potential TD
caused by different architecture design options, and then choose
the option with less potential TD. \\
\noindent \textbf{Life Cycle cost planning}: develop cost-effective plans that look
at the system throughout the life cycle to minimize overall TD of
the system. \\
\noindent \textbf{Human factors analysis}: cultivate a culture that minimizes the
unintentional TD caused by human factors (e.g., indifference
and ignorance).

\subsection{Discussion}
Managing Technical Debt is a difficult and a subjective task. Each one would have a different approach, We here used SonarQube, PMD (at times) and Code Analytix to perform TD Management activities. We learned that no one tool can get us holistic view of all the activities. Also, different tools shows up different results. Hence, there is no clear set guidelines as to when and which tool one should go for. Issue wise, we had issues while running SonarQube due to incompatible Java version. Another issue we faced was with the projects analyzed. As we didn't had much insights about the project, analyzing some TD items was difficult. \\
Using tools in combination, we overcome the very first issue. Also, the choice of the tools is purely subjective and project dependent decision. Referring to official documentation of SonarQube and Java helped to resolve all the technical issues faced while installing and analyzing projects. \\

This study turned out to be interesting as we got an insight about how TDM activities are conducted in a company for one or more projects. We explored some tools and got an understanding of what and how each tool serves role in
Technical debt management. Nevertheless, this project required
a lot of research about the tools, lot of time was utilized in
dealing with plugin based tools like PMD to get an overview of
the TD items it caters since it lacks representation of the
complete technical debt.

\section{Principal calculation model \& Tools}
\subsection{TD principal calculation model} 
\noindent There are mainly three approaches that estimate Technical
Debt Principal in a given system. out of these, we chose the
method supported by SonarQube TD plugin. Our TD Principal
focuses on the following based on the TD items we identified so
far: \\
\noindent \textbf{Duplication} : Estimated effort required to remove duplicates
from the code.\\
\noindent \textbf{Bugs} : Estimated effort to fix bug issues. \\
\noindent \textbf{Vulnerabilities} : Estimated time/effort to fix vulnerability
issues.\\
\noindent \textbf{Code smells} : Effort to fix all maintainability issues.
Comments : Estimated effort associated with documenting the
undocumented portions of the API. \\
\noindent \textbf{Coverage} : effort required to bring coverage from 0\% to 80\%. \\
\noindent  \textbf{Complexity} : total estimated effort needed to split every
method and every class (of those requiring such a split).\\
\noindent \textbf{Design}: estimated effort associated with cutting all existing
edges between files.\\
TD is summation of all the mentioned dimensions above.\\

\noindent Where, \\
Duplication = (cost to fix one block) * (duplicated
blocks)*(US\$ per hour)\\
\\
Violations = (cost to fix high severity violations * number of high severity violations + cost to fix medium severity violation * number of medium severity violations + cost to fix low severity violations * number of low severity violations) * (US\$
per hour)\\
\\
Comments = (cost to comment one API) * (public undocumented API) * (US\$ per hour)\\
\\ 
Coverage = (cost to cover uncovered lines of code) * (uncover
ed lines) * (US\$ per hour)\\
\\
Design=(cost to cut an edge between two files * package edges weight) * (US\$ per hour)\\
\\
Complexity = (cost to split a method) * (function complexity 
distribution $\geq$	 8) + (cost to split a class) * (class complexity 
distribution $\geq$	 60) \\
\\
Here, we are considering two types of complexities - cyclomatic
and cognitive complexity. While cyclomatic complexity
determines the difficulty of testing your code, cognitive
complexity determines the difficulty of reading and
understanding the code. \\
\\
Default Values for Parameters:
\begin{center}
\begin{table}[hbt!]
\centering
\caption{Default Values for Parameters}
\begin{tabular}{ |c|c|c|c| } 
\hline
Cost & Default Value (in dollars)  \\
\hline
cost to fix one block & 2 \\ 
\hline
cost to fix high severity violations & 0.5
 \\ 
\hline
cost to fix medium severity violations & 0.3 \\ 
\hline
cost to fix low severity violations & 0.1 \\ 
\hline
cost to comment one API & 0.2 \\ 
\hline
cost to cover uncovered lines of code &  0.2\\ 
\hline
cost to cut an edge between two files & 4 \\ 
\hline
cost to split a method & 0.5 \\ 
\hline
cost to split a class & 8 \\ 
\hline

\end{tabular}
\label{tab:PPer}
\end{table}
\end{center}

\subsection{More tools to manage Technical debt - CodePro Analytix} 
\noindent CodePro AnalytiX is a comprehensive automated software code
quality and security analysis tool which is tightly integrated into
Eclipse, it guarantees superior code quality, maximum
developer productivity and Project maintainability by adding its
potential features like code audit, metrics, testing and team
collaboration while giving continuous quality improvement
throughout the entire code development cycle. It is used by
companies to save time, money and manage technical debts \cite{b7}. \\

\noindent Features of CodePro Analytix:
Dynamic, extensible tools that detect, report and repair
instances of non-compliance with prede fined coding standards
and style conventions. It detects and corrects code quality issues
automatically. It helps in distribution of quality standards across
development team. It gives higher quality software product.
CodePro Analytix contains 960+ audit rules and matrices and
350+ quick fixes. It also allows customization of rules, Metrics
sets and rule sets. It provides dependency analysis and provides
report for the same. It generates multiple report forms (HTML,
XML, CSV). It contains Javadoc analysis and repair
functionality. Dependency Analysis: Analyzes dependencies
between projects, packages and classes \cite{b5}\cite{b7}\cite{b7} \\
\\
\noindent Installation steps for CodePro Analytix in Eclipse:\\
\noindent Step 1: Open Eclipse IDE - Goto Help - Install new
Software \\
\begin{figure}[hbt!]
\centering
    \frame{\includegraphics[width=\columnwidth]{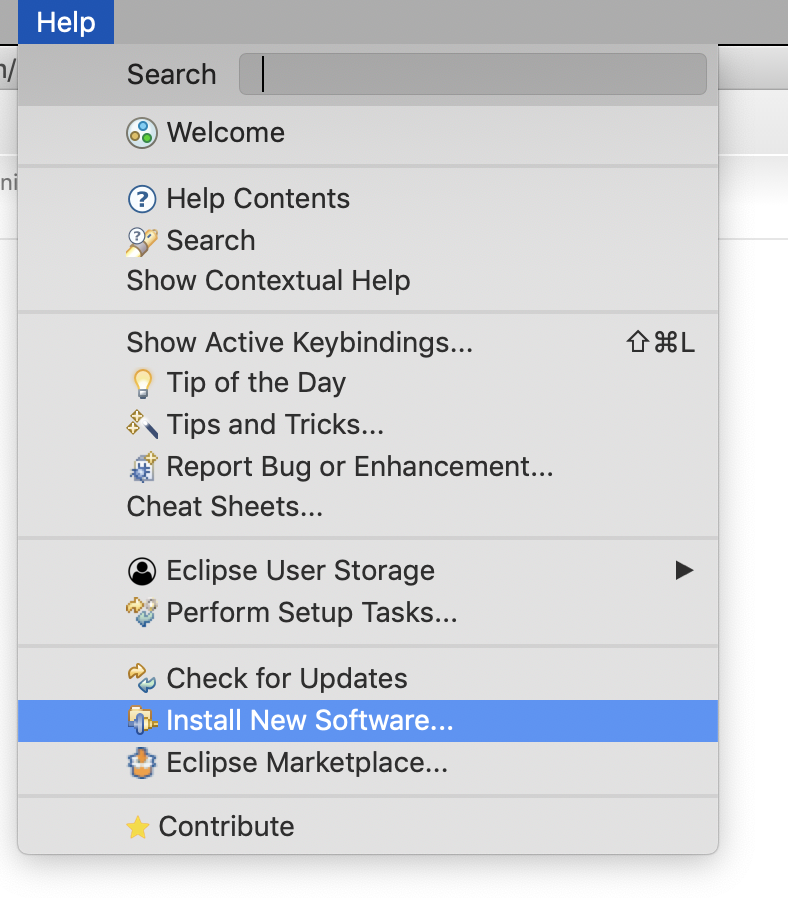}}
    \caption{Step 1.} 
\end{figure}
\noindent Step 2: Enter the URL and then click on Add button. Select CodePro from the options \\
\begin{figure}[hbt!]
\centering
    \frame{\includegraphics[width=\columnwidth]{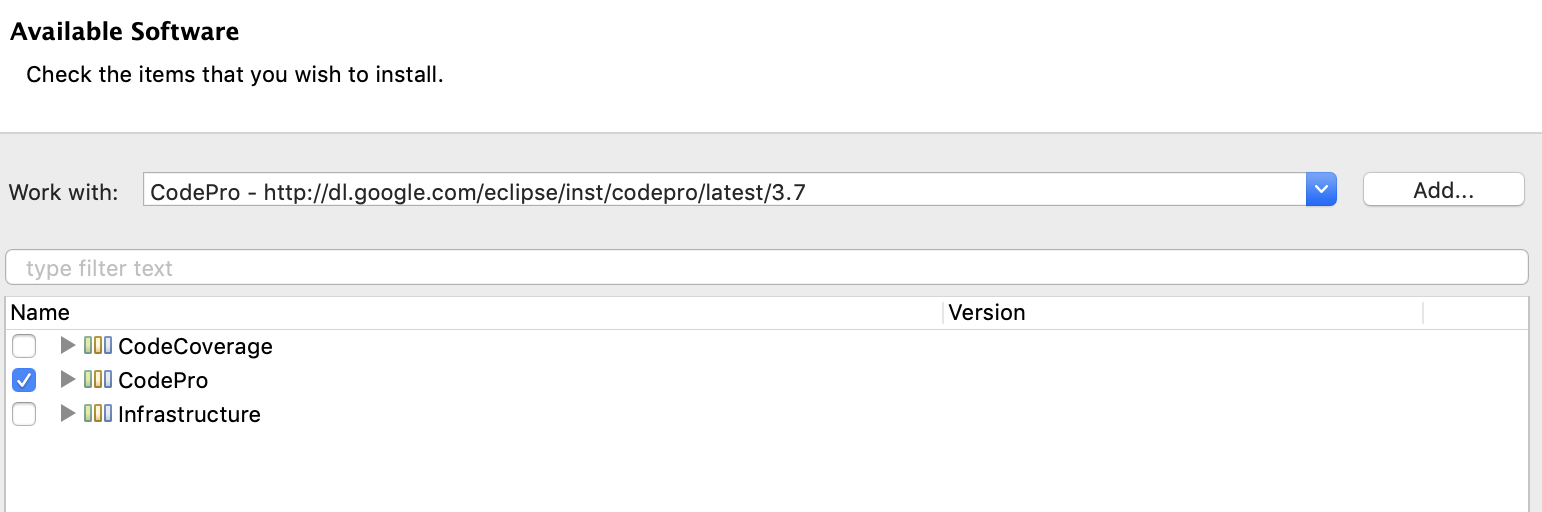}}
    \caption{Step 2.} 
\end{figure}
\noindent Step 3: Accept the terms and finish the installation. \\
\\
\noindent Analysis of project by using CodePro Analytix: \\
Code analysis is very important feature of CodePro Analytix,
which can be performed through the code auditing feature.
There are over 770 java-based coding rules in more than 30
categories built into a tool. Audit run for this area and determine
the location where the code has a problem. Running code audit
on all modules show some interesting code violations. The
Audit View shows the explanation and recommendation for
each violation as well \cite{b7}.
\begin{figure}[hbt]
\centering
    \frame{\includegraphics[width=\columnwidth]{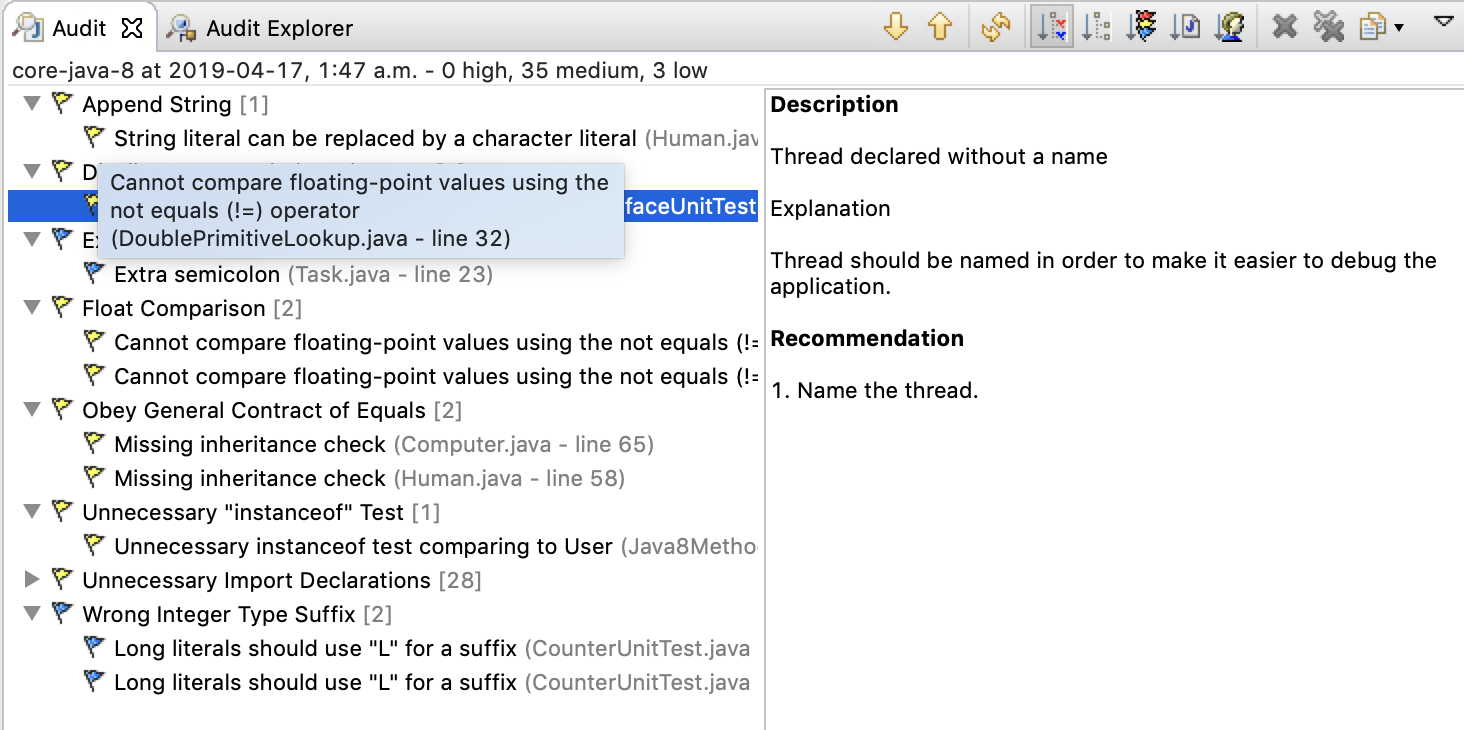}}
    \caption{Audit View - Project 1.} 
\end{figure}
\begin{figure}[hbt]
\centering
    \frame{\includegraphics[width=\columnwidth]{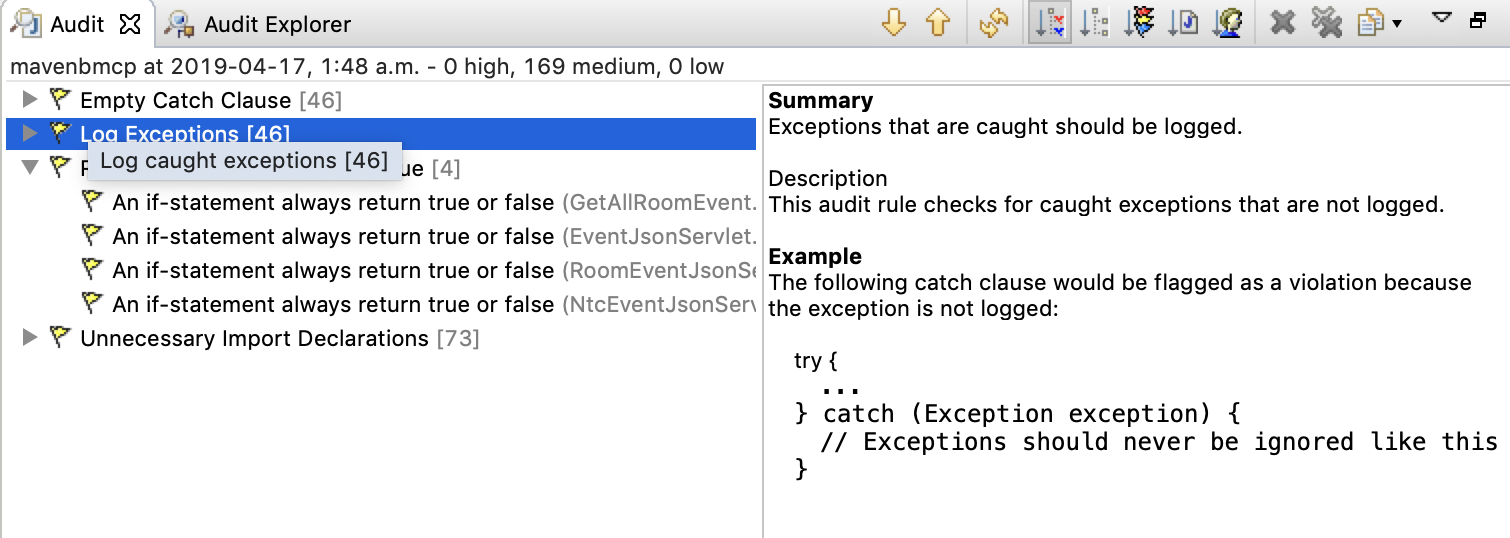}}
    \caption{Audit View - Project 2.} 
\end{figure}
\begin{figure}[hbt]
\centering
    \frame{\includegraphics[width=\columnwidth]{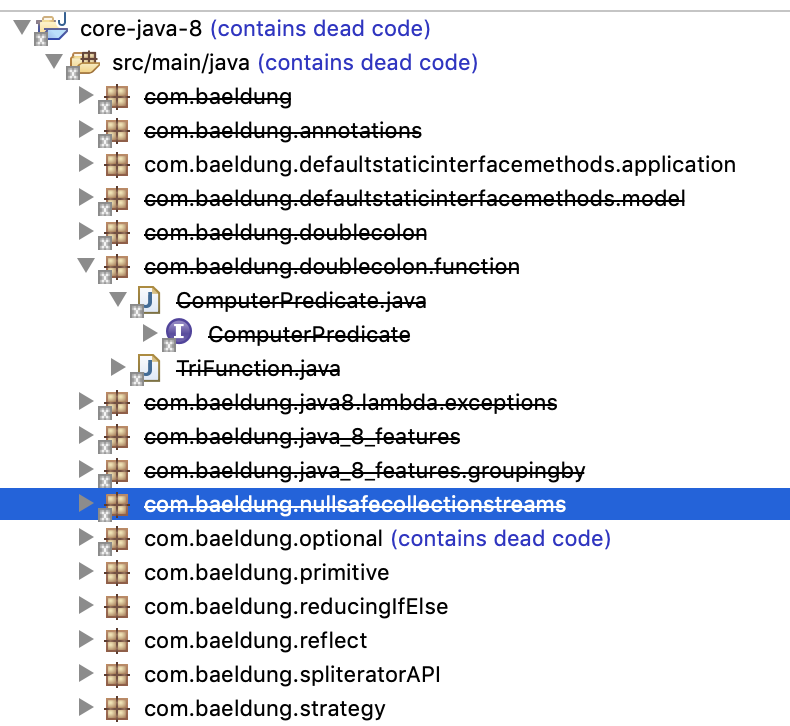}}
    \caption{Dead Code - Project 1.} 
\end{figure}
\begin{figure}[hbt]
\centering
    \frame{\includegraphics[width=\columnwidth]{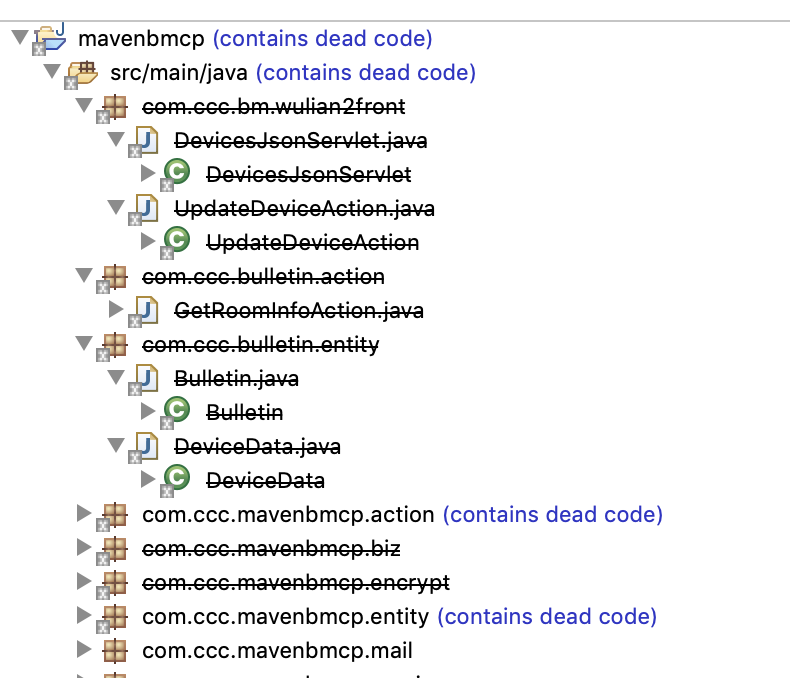}}
    \caption{Dead Code - Project 2.} 
\end{figure}
\begin{figure}[hbt!]
\centering
    \frame{\includegraphics[width=\columnwidth]{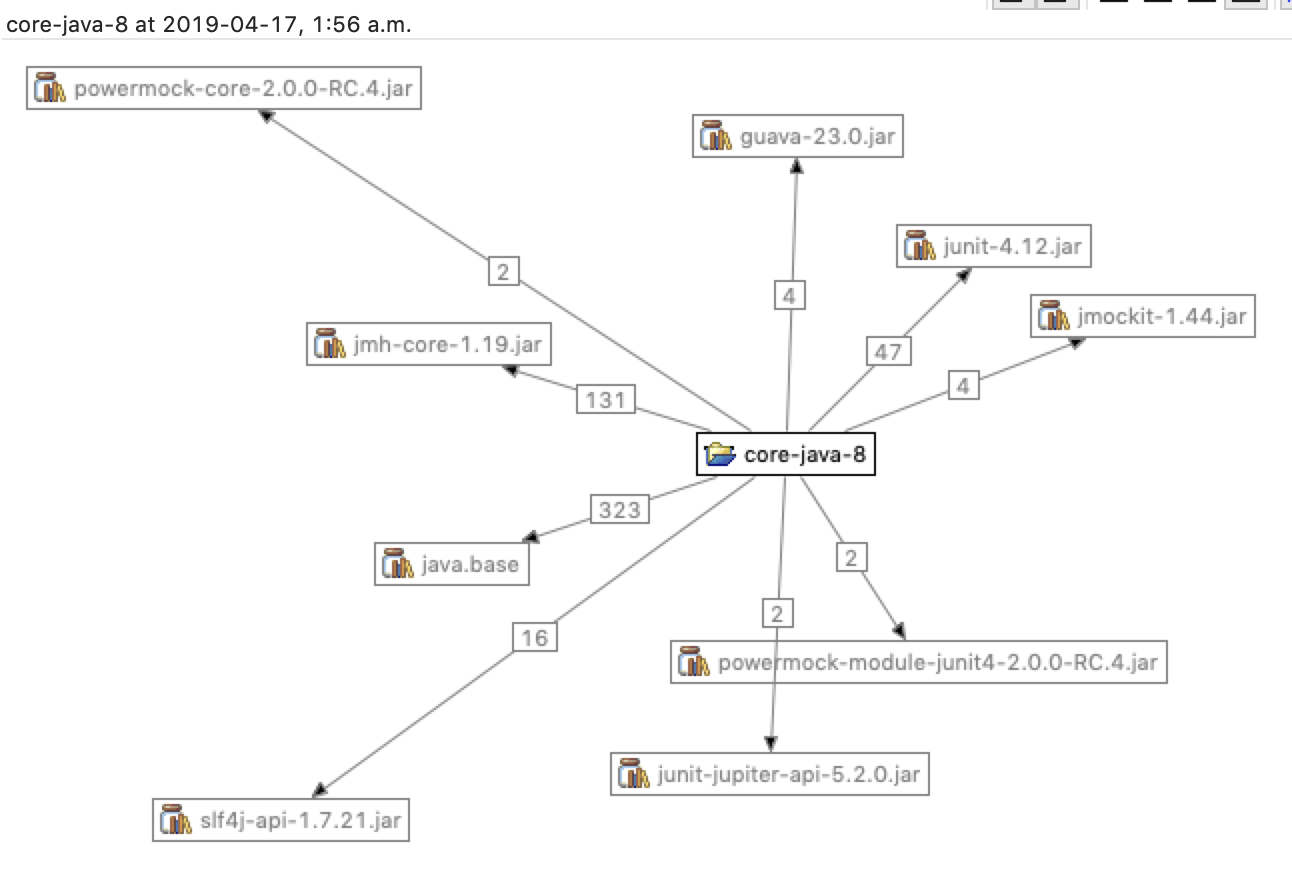}}
    \caption{Dependency Analysis - Project 1.} 
\end{figure}
\begin{figure}[hbt!]
\centering
    \frame{\includegraphics[width=\columnwidth]{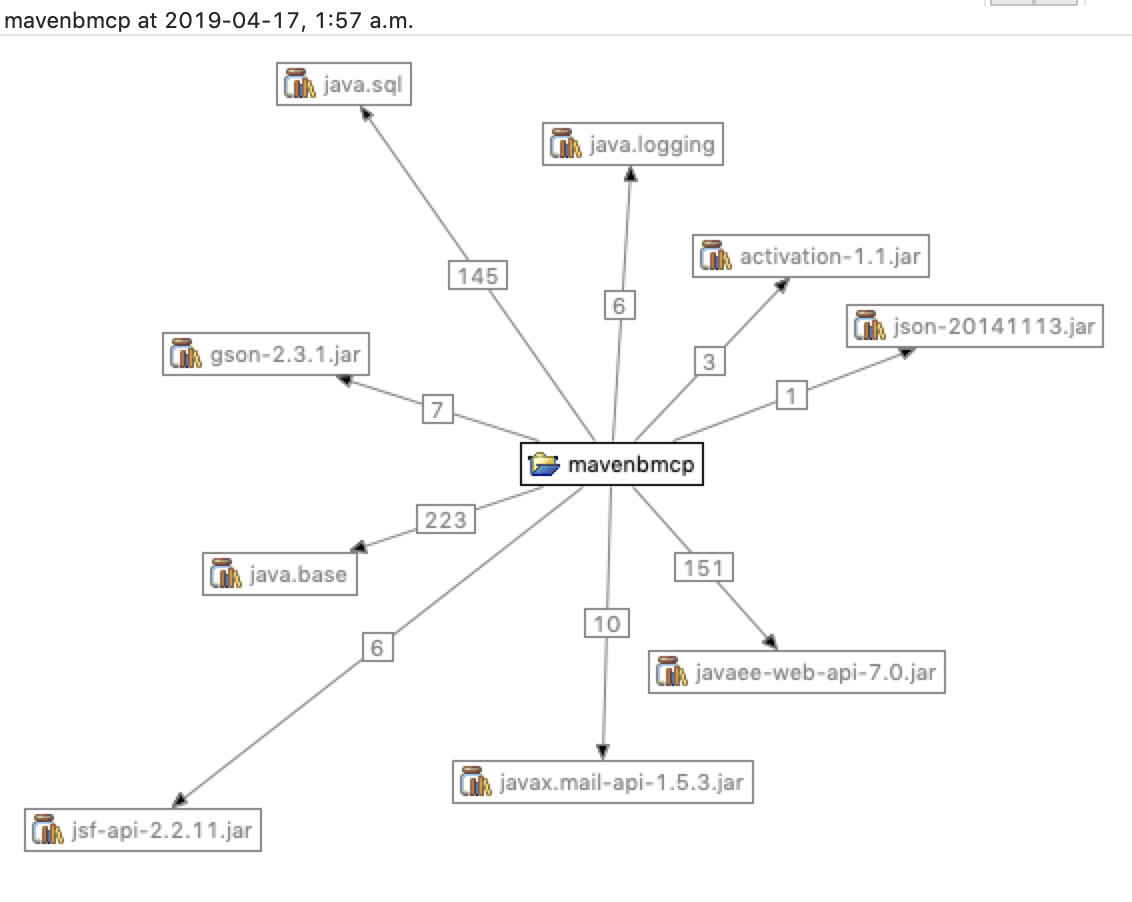}}
    \caption{Dependency Analysis - Project 2.} 
\end{figure}
\begin{figure}[hbt!]
\centering
    \frame{\includegraphics[width=\columnwidth]{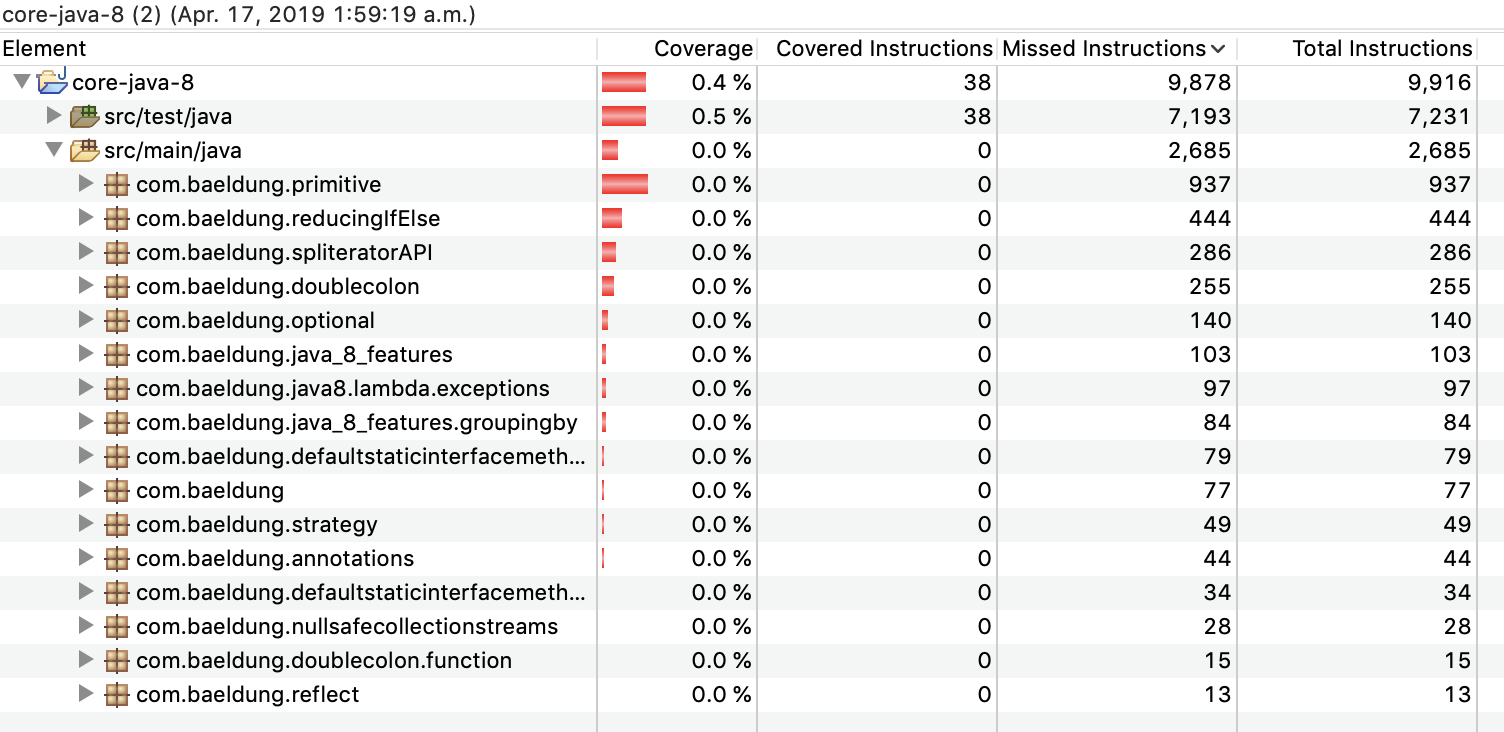}}
    \caption{Code Coverage - Project 1.} 
\end{figure}
\\
\noindent Please note, there are no JUnit coverage in Project 2.
\section*{Conclusion}
\noindent We studied SonarQube in detail and applied the tool on our chosen projects. To start with, we analyzed various quality attributes like reliability, maintainability
and security. Next, we studied all the TD Management activities such as Identification, Measuring, Monitoring, Repayment and Prevention. We used SonarQube to perform each of this activity in depth. In addition to the given tools we also studied CodePro Analytix and gained more insights. Finally, the tools we used are
found to be effective and give important information about
technical debt. There are a lot of advantages of these tools but
still there are few limitations. We can say no tool is perfect in finding all kinds of technical debts. Also, our work didn't include environmental and requirement debts.

\clearpage
\section*{Appendix}

\noindent \textbf{Technical Debt Identification}\\
\noindent Project 1: Core Java 8 - Test Debt \\
\begin{figure}[hbt!]
\centering
    \frame{\includegraphics[width=\columnwidth]{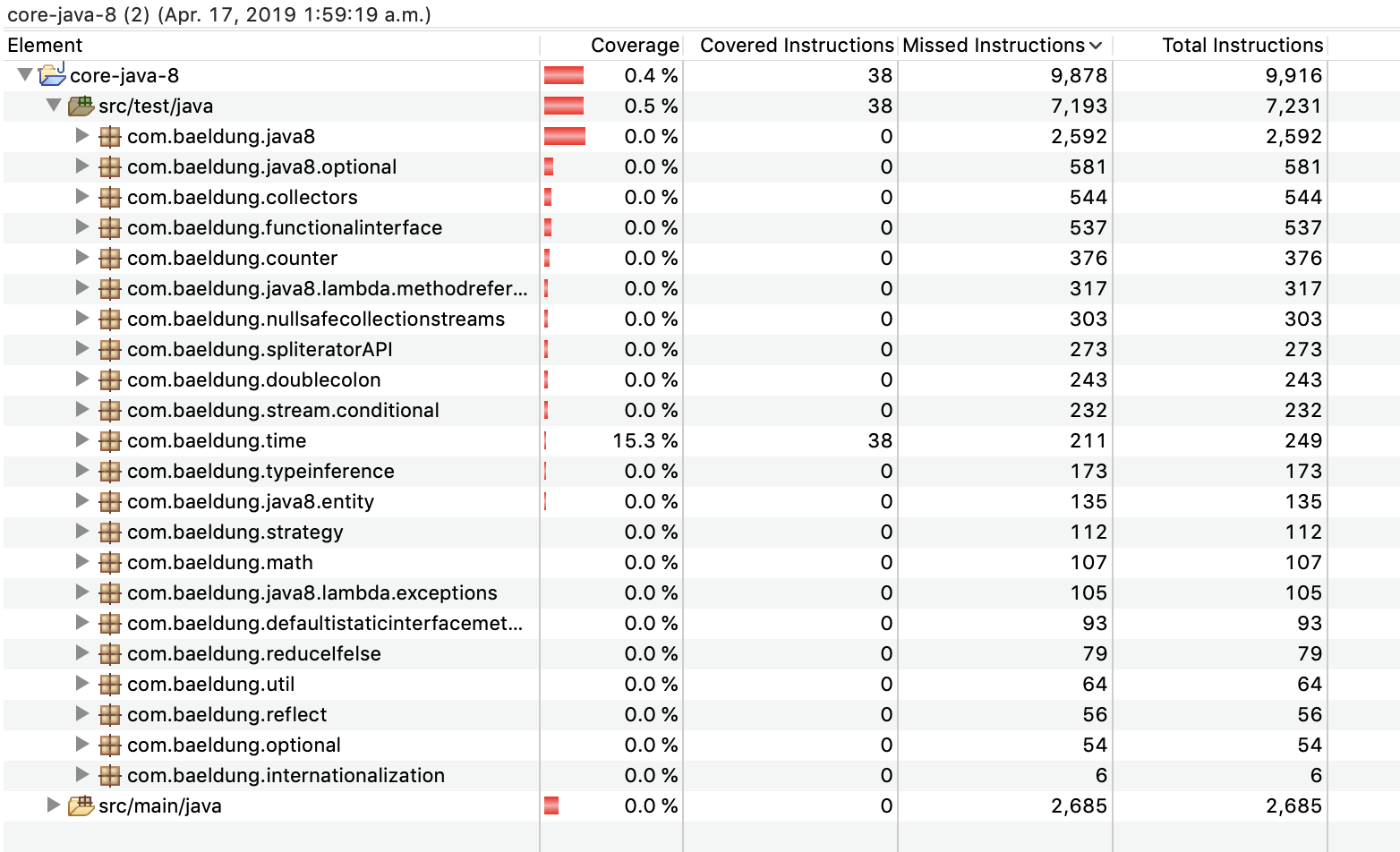}}
    \caption{Project 1: Core Java 8 - Test Debt.} 
\end{figure}

\noindent Project 1: Core Java 8 - Violation Overview using PMD \\
\begin{figure}[hbt!]
\centering
    \frame{\includegraphics[width=\columnwidth]{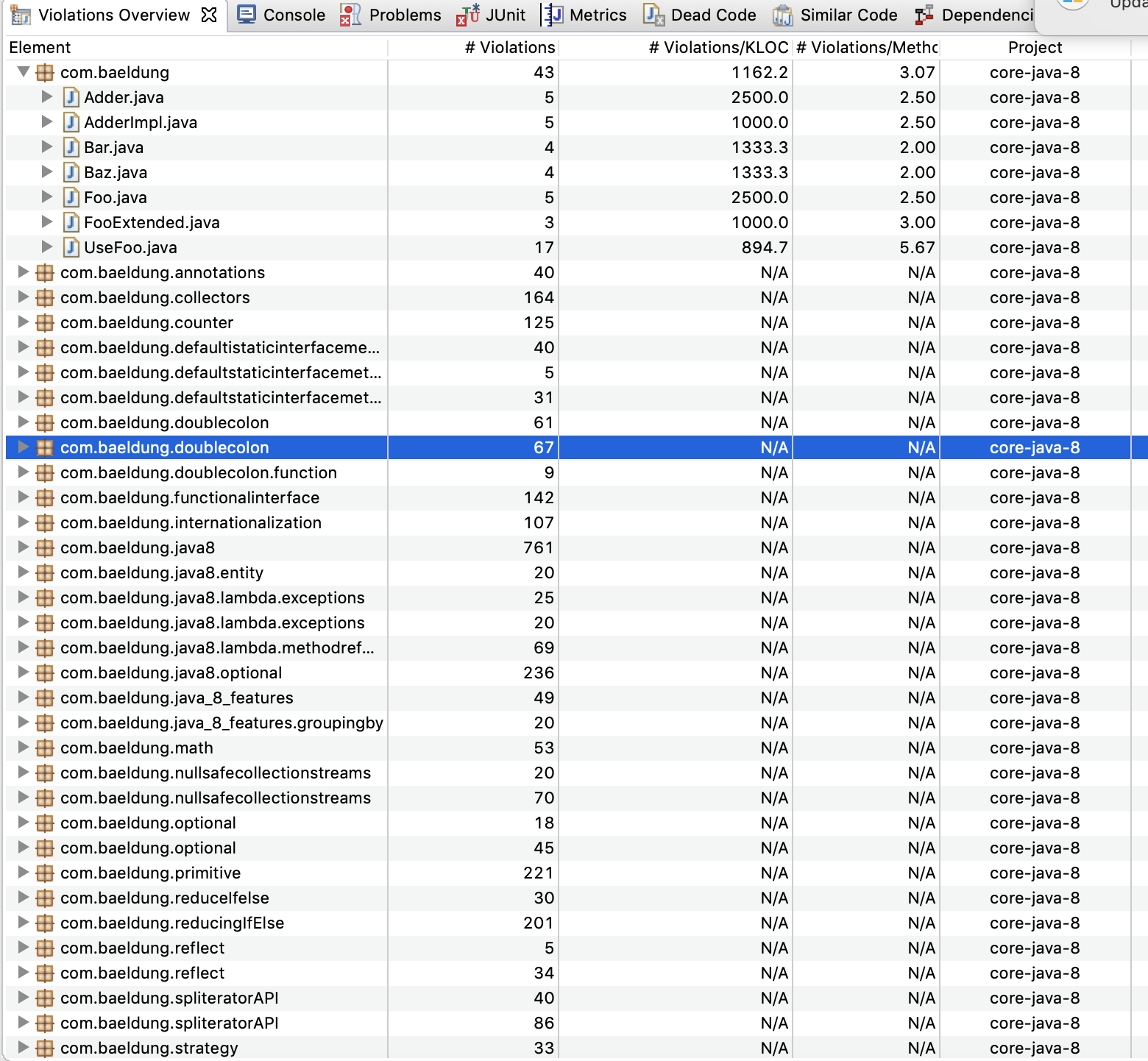}}
    \caption{Project 1: Core Java 8 - Violation Overview using PMD.} 
\end{figure}
\noindent Project 2: Booking Manager - Violation Overview using PMD \\
\begin{figure}[hbt!]
\centering
    \frame{\includegraphics[width=\columnwidth]{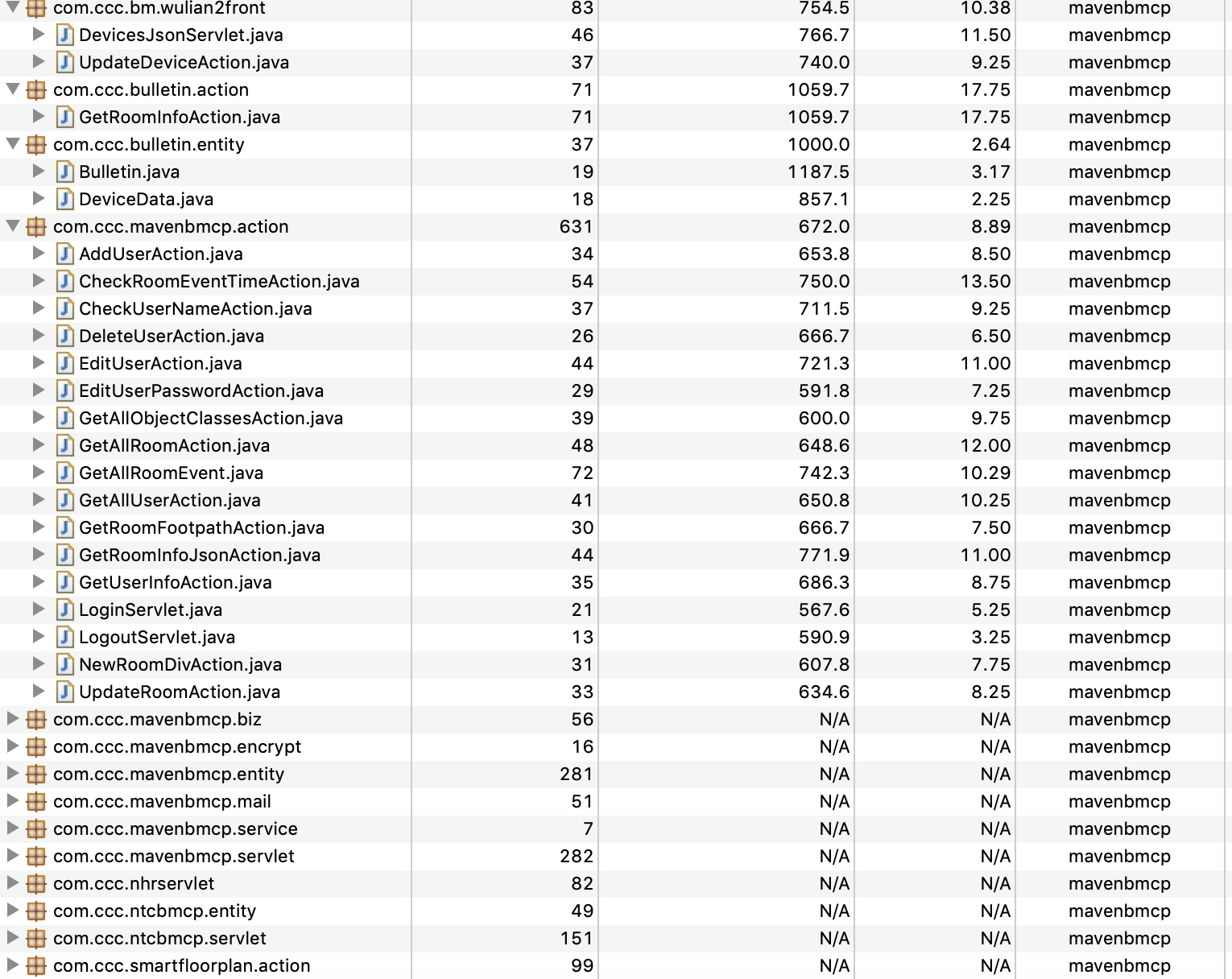}}
    \caption{Project 2: Booking Manager - Violation Overview using PMD.} 
\end{figure}

\noindent \textbf{TD Representation} \\
\begin{center}
\begin{table}[hbt!]
\centering
\caption{TD Representation - Project 1}
\begin{tabular}{ |c|c|c|c| } 
\hline
ID & 1.4  \\
\hline
Name & Code Smell \\ 
\hline
Location &  src/../doublecolon/ComputerUtils.java \\ 
\hline
Responsible/author & Not Assigned \\ 
\hline
Dimension & Code Debt \\ 
\hline
Date/time & Apr 14, 2019 15 : 37 :18 \\ 
\hline
Context &  Utility classes should not have public constructors \\ 
\hline
Propagation rule & No impact to other classes \\ 
\hline
Intentionality & Unintenstional\\ 
\hline
\end{tabular}
\label{tab:PPer}
\end{table}
\end{center}
\begin{center}
\begin{table}[hbt!]
\centering
\caption{TD Representation - Project 1}
\begin{tabular}{ |c|c|c|c| } 
\hline
ID & 1.5  \\
\hline
Name & Code Smell \\ 
\hline
Location & src/.../doublecolon/MacbookPro.java \\ 
\hline
Responsible/author & Not Assigned \\ 
\hline
Dimension & Code Debt \\ 
\hline
Date/time & Apr 14, 2019 15 : 37 :18 \\ 
\hline
Context &  "Preconditions" and logging arguments \\ 
\hline
Propagation rule & No impact to other classes \\ 
\hline
Intentionality & Unintenstional\\ 
\hline
\end{tabular}
\label{tab:PPer}
\end{table}
\end{center}

\begin{center}
\begin{table}[hbt!]
\centering
\caption{TD Representation - Project 1}
\begin{tabular}{ |c|c|c|c| } 
\hline
ID & 1.6  \\
\hline
Name & Code Smell \\ 
\hline
Location & src/.../LambdaExceptionWrappers.java \\ 
\hline
Responsible/author & Not Assigned \\ 
\hline
Dimension & Code Debt \\ 
\hline
Date/time & Apr 14, 2019 15 : 37 :18 \\ 
\hline
Context &  Generic exceptions \\ 
\hline
Propagation rule & No impact to other classes \\ 
\hline
Intentionality & Unintenstional\\ 
\hline
\end{tabular}
\label{tab:PPer}
\end{table}
\end{center}

\end{document}